\documentclass[preprintnumbers,amsmath,amssymb,floatfix,12pt,prd,superscriptaddress,nofootinbib]{revtex4}

\usepackage{graphicx}
\usepackage{epsfig}
\usepackage{bm}
\usepackage{amsfonts}
\usepackage{subfigure}
\usepackage{multirow}
\usepackage{color}
\usepackage{float}
\usepackage{xfrac}
\usepackage{newtxtext,newtxmath}
\usepackage{xcolor}

\begin{document}

\title{\textbf{Structure formation in a non-canonical scalar field model of clustering\\ dark energy}}

\author{Zanyar Ebrahimi}
\email{zebrahimi@maragheh.ac.ir}\affiliation{Research Institute for Astronomy and Astrophysics of Maragha, University of Maragheh, Maragheh, P.O.Box 55136–553, Iran}

\author{Kayoomars Karami}
\email{kkarami@uok.ac.ir}\affiliation{Department of Physics, University of Kurdistan, Pasdaran Street, P.O. Box 66177-15175, Sanandaj, Iran}

\date{\today}

\begin{abstract}
This paper examines the growth of dark matter and dark energy perturbations within a non-canonical scalar field model characterized by an exponential potential. Through dynamical system analysis, we identify critical points and track the background evolution of a spatially flat FLRW universe dominated by dark energy and pressureless dark matter. We systematically derive key cosmological quantities, including the Hubble parameter, deceleration parameter, density parameters, and the scalar field's equation of state, and explore their dependence on model parameters. Within the linear perturbation framework, employing the pseudo-Newtonian formalism, we compute the growth factor of matter density perturbations. To investigate the non-linear regime of structure formation, we employ the spherical collapse model and derive its key parameters. Building on these findings, we compute the function $f(z)\sigma_8(z)$ and the relative number density of halo objects exceeding a given mass threshold. Our results indicate that non-canonical scalar field models can effectively account for both background cosmic evolution and the growth of structure, offering potential insights into observational constraints and large-scale dynamics.
 \end{abstract}


\maketitle

\newpage
\section{Introduction}

The accelerating expansion of the Universe stands as one of the most intriguing and profound mysteries in modern cosmology. Over the past two decades, an abundance of observational data has pointed towards this remarkable phenomenon. Evidence from studies on type-Ia supernovae \citep{Riess1998, Perlmutter1999, Kowalski2008} \citep[for a review see][]{Goobar2011}, the cosmic microwave background (CMB) \citep{Komatsu2009, Jarosik2011, Komatsu2011, Planck2016, Planck2020}, measurements of the Hubble constant and large-scale structures (LSS)\citep{Cole2005}, baryonic acoustic oscillations \citep{Tegmark2004, Cole2005, Eisenstein2005, Percival2010, Blake2011, Reid2012, Alam2017}, high-redshift galaxies \citep{Alcaniz2004}, high-redshift galaxy clusters \citep{Wang1998, Allen2004},
and weak gravitational lensing \citep{Benjamin2007, Amendola2008, Fu2008} have collectively supported the notion of an accelerating universe.

This observational framework, when combined with the principles of general relativity on cosmic scales, has led researchers to propose two broad theoretical approaches: the introduction of dark energy (DE) and modifications to the theory of gravity. The standard paradigm within Einstein's general relativity assumes the existence of an unknown exotic component with negative pressure --- dark energy --- that drives cosmic acceleration. Dark energy, characterized by an equation-of-state (EoS) parameter $\omega<-1/3$, emerges as a compelling theoretical construct capable of providing a viable explanation for the observed cosmic acceleration. On the other hand, modified gravity (MG) theories propose that the observed acceleration can be attributed to fundamental changes in the laws of gravity rather than an additional energy component \citep[for a good review on MG see][]{Shankaranarayanan2022}. These two approaches continue to be the subject of active research, providing competing explanations for the universe’s accelerating expansion.

The standard $\Lambda$CDM model, which integrates a cosmological constant $\Lambda$ with an equation of state $\omega = -1$ as a candidate for dark energy \citep{Peebles2003}, alongside cold dark matter (CDM), has been successful in providing a framework that explains many observational features of the Universe, such as the cosmic acceleration, the cosmic microwave background (CMB) and large-scale structures. However, this model faces significant theoretical challenges, including issues of fine-tuning and the cosmic coincidence problem \citep{Weinberg1989, Sahni2000, Carroll2001, Padmanabhan2003, Copeland2006}. To overcome these challenges, alternative models for dark energy with a dynamic EoS have been proposed. These models encompass various classes, including quintessence \citep{Caldwell1998, Erickson2002}, phantom \citep{Caldwell2002, Caldwell2003, Elizalde2004}, K-essence \citep{Armendariz2000, Armendariz2001, Chiba2000}, chaplygin gas \citep{Kamenshchik2001}, generalized chaplygin gas \citep{Bento2002}, dilaton \citep{Gasperini2001, Arkani2004, Piazza2004}, agegraphic \citep{Cai2007}, new agegraphic \citep{Wei2008}, and quintom \citep{Wei2005} which involve scalar fields that possess the potential to elucidate the elusive nature of dark energy and offer intriguing avenues for understanding the acceleration of the Universe without the need for a cosmological constant.

Furthermore, extensive research has been conducted on the formation and growth of structures predicted by both approaches of modified gravity \citep{Koyama2006, Malekjani2009, Brax2012, Nesseris2013, Asadzadeh2016, Nazari-Pooya2016, Malekjani2017, Nunes2018} and dark energy \citep[see e.g.][among many others]{Linder2003, abramo2007, Batista2013, Malekjani2015, Naderi2015, Rezaei2017, fahimi2018, Rezazadeh2020, Ziaie2020}. These studies have highlighted the significant role of large-scale structure in distinguishing between different DE and MG models. The intricate interplay between the expansion history and structure formation underscores the importance of investigating large-scale structure to unravel the mysteries surrounding dark energy and modified gravity.

It is widely theorized that cosmic structures originated from the gravitational collapse of primordial density fluctuations, which are believed to have emerged during the inflationary epoch \citep{Guth1981, Linde1990}. During this phase, quantum fluctuations were rapidly stretched to macroscopic scales by inflation, imprinting the initial perturbations onto the matter distribution \citep{Mukhanov1981, Starobinsky1982}. Over time, these perturbations grew under gravitational attraction, giving rise to vast cosmic structures, including galaxies and galaxy clusters \citep{Peebles1993, Peacock1999}. In the early universe, density fluctuations were sufficiently small to be treated within the linear regime, where their evolution could be described using perturbative methods \citep{abramo2008}. As perturbations strengthened, they entered the nonlinear regime, requiring more advanced models such as the spherical collapse model to track their progression beyond linear theory \citep{Gunn1972, Press1974}. Both regimes (linear and nonlinear) can be effectively studied using the pseudo-Newtonian formalism, which provides a consistent framework for analyzing the growth of structure in an expanding universe. The complex interplay among dark matter, dark energy, and baryonic interactions shaped the dynamic of the formation of cosmic structures, leaving an observable imprint on the large-scale universe \citep{Peebles2003, Bromm2011}.

Canonical scalar field models, such as quintessence and phantom dark energy, provide dynamical alternatives to the cosmological constant by introducing scalar fields with standard kinetic terms. Quintessence, governed by a slowly rolling scalar field, exhibits an evolving equation of state distinct from $\Lambda$CDM \citep{Peebles2003}. Phantom models, characterized by a negative kinetic term, lead to super-accelerated expansion with $\omega<-1$, potentially culminating in a Big Rip scenario \citep{Caldwell2002}. In these models, dark energy (DE) perturbations are typically negligible on small scales due to their high effective sound speed ($c_{\rm eff}\sim 1$), which suppresses growth inside the sound horizon \citep{Caldwell1998, Erickson2002}. By contrast, non-canonical scalar field models, such as k-essence, introduce modified kinetic terms that allow DE perturbations to propagate with a much lower effective sound speed. This enables clustering on sub-horizon scales, providing a richer structure formation dynamic than standard quintessence models \citep{Armendariz2000, Chiba2000, Armendariz2001}. The flexibility of non-canonical models in addressing fine-tuning issues and accommodating observational constraints makes them compelling alternatives to traditional scalar field approaches.

This study focuses on k-essence, a non-canonical scalar field model derived from the well-known Horndeski Lagrangian, which encompasses higher-order derivatives of the field. As the simplest subset of this framework, k-essence generalizes canonical scalar field models by allowing its Lagrangian to depend not only on the scalar field itself but also on its kinetic term \citep{Armendariz2000}. A key distinction of k-essence is that, unlike canonical models, its kinetic energy can independently drive cosmic acceleration, reducing reliance on potential-driven dynamics \citep{Chiba2000}. Initially proposed for the early inflationary epoch \citep{Garriga1999}, k-essence was later adapted to describe dark energy \citep{Chiba2000}. By considering power-law kinetic terms, this work examines the linear and non-linear growth of fluctuations in k-essence scenarios and their implications for cosmic structure formation, providing alternative perspectives on the evolution of large-scale structures.

The paper is structured as follows: In Section 2, we present the theoretical framework of the non-canonical scalar field model, highlighting its Lagrangian formulation. In Section 3, we introduce a dynamical system analysis to solve the equations of motion and derive the background evolution of the universe. Section 4 is dedicated to analyzing how the model parameters influence the initial conditions and shape the evolution of key background quantities. Section 5 focuses on the linear perturbation theory, examining the growth of fluctuations and their implications for cosmic structure formation. Next, in Section 6, we study the spherical collapse of perturbations in the non-linear regime. Finally, in Section 7, we summarize the key findings and conclude the paper.


\section{Non-Canonical Scalar Field Model}

Within the framework of Einstein gravity, the action of the non-canonical scalar field $\phi$ minimally coupled to the gravity is given by
\begin{equation}\label{action}
	S=\int{d^4 x \sqrt{-g}\left(\frac{M_{\rm P}^2}{2}R+\mathcal{L}(X,\phi)\right)}+S_m,
\end{equation}
where, $g$ is the determinant of the background metric, $M_{p}=(8\pi G)^{-1/2}$ is the reduced Planck mass, and $R$ is the Ricci scalar. In Eq. (\ref{action}), $S_m$ is the action of matter and $\mathcal{L}(X,\phi)$ is the Lagrangian of a non-canonical scalar field defined as follows
\begin{equation}\label{lagrangian}
	\mathcal{L}(X,\phi)=X\left(\frac{X}{M^4}\right)^{\alpha-1}-V(\phi),
\end{equation}
where, $\alpha$ and $M$ are constants. In Eq. (\ref{lagrangian}) $X\equiv (1/2)g^{\mu\nu}\partial_\mu \phi\partial_\nu \phi$ is the canonical kinetic term and $V(\phi)$ is a self-interacting potential for the scalar field. Note that in the case of $\alpha=1$, Eq. (\ref{lagrangian}) reduces to the Lagrangian of the standard canonical scalar field.

The energy density $\rho_\phi$ and pressure $p_\phi$ of the scalar filed in terms of the Lagrangian (\ref{lagrangian}) are given by
\begin{align}
	\rho_\phi&=2X\frac{\partial \mathcal{L}(X,\phi)}{\partial X}-\mathcal{L}(X,\phi),\label{rhophi}\\
	p_\phi&=\mathcal{L}(X,\phi). \label{pphi}
\end{align}

We consider a spatially flat Friedmann-Lima\^{\i}tre-Robertson-Walker (FLRW) universe containing pressure-less matter and clustering dark energy (DE) with the line element
\begin{equation}\label{metric}
	ds^2=dt^2-a^2(t)\left(dr^2+r^2 d\Omega^2 \right).
\end{equation}
Variation of the action (\ref{action}) with respect to the metric (\ref{metric}) gives the Friedmann equations as follows
\begin{align}
	&H^2=\frac{1}{3M_P^2}\left((2\alpha-1)X\left(\frac{X}{M^4}\right)^{\alpha-1}+V(\phi)+\rho_m\right),\label{Fr1}\\
	&\dot{H}=-\frac{1}{2M_p^2}\left(2\alpha X\left(\frac{X}{M^4}\right)^{\alpha-1}+\rho_m\right), \label{Fr2}
\end{align}
where the dot sign denotes a derivative with respect to the cosmic time $t$, and $H\equiv \dot{a}(t)/a(t)$ is the Hubble parameter with $a(t)$ being the scale factor of the universe.  Here, $\rho_m$ is the matter energy density. The canonical kinetic term $X$ for the spatially homogeneous metric (\ref{metric}) reduces to $X=\dot{\phi}^2/2$.

Variation of the action (\ref{action}) with respect to $\phi$ gives the equation governing the scalar field as follows
\begin{equation}\label{phieq}
	\ddot{\phi}+\frac{3H\dot{\phi}}{2\alpha-1}+\frac{V_{,\phi}}{\alpha(2\alpha-1)}\left(\frac{2M^4}{\dot{\phi}^2}\right)^{\alpha-1}=0,
\end{equation}
where the subscript "$,\phi$" denotes a derivative with respective to $\phi$. The continuity equations governing the pressure-less ($p_m=0$) matter and the scalar field, respectively, are given by
\begin{align}
	&\dot{\rho}_m+3H\rho_m=0,\label{rhomcon}\\
	&\dot{\rho}_\phi+3H(\rho_\phi+p_\phi)=0. \label{rhophicon}
\end{align}
Note that one can obtain Eq. (\ref{phieq}) by inserting Eqs. (\ref{lagrangian})-(\ref{pphi}) into (\ref{rhophicon}). In the context of a non-canonical scalar field, the speed of sound that represents the propagation speed of the DE fluctuations with respect to the uniform background is obtained as follows
\begin{equation}
	C_s^2\equiv\frac{\partial p_\phi}{\partial \rho_\phi}=\left(\frac{\partial p_\phi}{\partial X}\right)/\left(\frac{\partial \rho_\phi}{\partial X}\right)=\frac{1}{2\alpha-1}.
\end{equation}
To prevent ghost instability and adhere to the subluminal condition, the speed of sound must fall within the range $0<C_s^2 \leqslant 1$. This places a constraint on the parameter $\alpha$ such that $\alpha\geqslant1$.

\section{Dynamical System Analysis}
In this section, we solve the set of cosmological equations by expressing them as an autonomous system of ordinary differential equations. To do this, we introduce the following set of dimensionless phase space variables
\begin{align}\label{dynvar}
	&x\equiv \frac{\sqrt{\rho_X}}{\sqrt{3}M_P H},~~~~y\equiv \frac{\sqrt{V}}{\sqrt{3}M_P H},~~~~\sigma\equiv -\frac{M_P}{\sqrt{3|\rho_X|}}\frac{{\rm d}\ln V}{{\rm d} t},
\end{align}
where
\begin{equation}\label{rhox}
	\rho_X\equiv\rho_\phi-V=(2\alpha-1)X\left(\frac{X}{M^4}\right)^{\alpha-1},
\end{equation}
is the kinetic part of the energy density associated with the scalar field. Taking derivatives of $x$, $y$ and $\sigma$ with respect to the $e$-folds variable $N=\ln a$ and using Eqs. (\ref{Fr1})-(\ref{phieq}) and (\ref{dynvar}), the evolution equations for $x$, $y$ and $\sigma$ are obtained as follows \citep{DeSantiago2013}
\begin{align}
	&\frac{{\rm d}x}{{\rm d}N}=\frac{3}{2}\left(\sigma y^2-x\frac{2\alpha}{2\alpha-1}\right)+\frac{3}{2}x\left(1-y^2+\frac{x^2}{2\alpha-1}\right),\label{dyn1}\\
	&\frac{{\rm d}y}{{\rm d}N}=-\frac{3}{2}x y \sigma +\frac{3}{2}y\left(1-y^2+\frac{x^2}{2\alpha-1}\right),\label{dyn2}\\
	&\frac{{\rm d}\sigma}{{\rm d}N}=-3\sigma^2 x(\Gamma-1)+\frac{3}{2}\sigma\frac{\alpha-1}{\alpha}\left(\frac{2\alpha}{2\alpha-1}-\sigma\frac{y^2}{x}\right),\label{dyn3}
\end{align}
where
\begin{equation}\label{Gamma}
	\Gamma\equiv\frac{V V_{,\phi\phi}}{V_{,\phi}^2}.
\end{equation}
Differentiating Eq. (\ref{Gamma}) with respect to $N$ and using the definitions of $x$ and $\sigma$ in Eq. (\ref{dynvar}) gives the evolution equation of $\Gamma$ as follows
\begin{equation}
	\frac{{\rm d}\Gamma}{{\rm d}N}=3 x \sigma\Big(\Gamma(\Gamma-1)-\Theta\Big),
\end{equation}
where
\begin{equation}\label{Theta}
	\Theta\equiv\frac{V^2 V_{,\phi\phi\phi}}{V_{,\phi}^3}.
\end{equation}
Here, we see that the new variable $\Theta$ is defined in terms of the third-order derivative of the potential with respect to $\phi$. If we continue this process and obtain the evolution equation of $\Theta$, a new variable proportional to the fourth-order derivative of $V$ with respect to $\phi$ appears in the equation. To truncate this series of equations, we may choose to constrain the function $V$ such that it gives constant values for $\Gamma$ or $\Theta$. For example, for $V(\phi)\propto \exp(-\lambda \phi/M_P)$ and $V(\phi)\propto\phi^n$ we have $\Gamma=1$ and $\Gamma=(n-1)/n$, respectively. Hence, for these potentials, the set of equations (\ref{dyn1})-(\ref{dyn3}) constitutes a closed system since the variable $\Gamma$ has a constant value. In the case of $\alpha=1$, the set of Eqs. (\ref{dyn1})-(\ref{dyn3}) reduces to the dynamical equations of the standard quintessence scalar field \citep[see e.g.][among many others]{Copeland1998, Caldwell1998, Zlatev1999, Macorra2000, Ng2001, Corasaniti2003, Caldwell2005, Linder2006, Yang2019, Joshi2021}. For a good review on the quintessence model, see \cite{Tsujikawa2013}.

In terms of the new variables (\ref{dynvar}), the density parameter and the equation of state (EoS) of the scalar field, the effective EoS, and the deceleration parameter  are obtained respectively as follows
\begin{align}
	&\Omega_\phi\equiv\frac{\rho_\phi}{3M_P^2 H^2}=x^2+y^2, \label{Ophi}\\
	&\omega_\phi\equiv\frac{p_\phi}{\rho_\phi}=\frac{x^2/(2\alpha-1)-y^2}{x^2+y^2},\label{ophi}\\
	&\omega_{\rm eff}\equiv\frac{p_\phi+p_m}{\rho_\phi+\rho_m}=\frac{x^2}{2\alpha-1}-y^2,\label{oeff}\\
	& q\equiv-1-\frac{\dot{H}}{H^2}=\frac{3}{2}\left(\frac{x^2}{2\alpha-1}-y^2\right)+\frac{1}{2}.\label{dec}
\end{align}
In order to express the Hubble parameter in terms of the dynamical variables, first we obtain the factor $\dot{H}/H^2$ by dividing Eq. (\ref{Fr2}) by $H^2$ and applying Eqs. (\ref{pphi}), (\ref{Fr1}), (\ref{Ophi}) and (\ref{ophi}) as follows
\begin{equation}\label{hdoth2}
	\frac{\dot{H}}{H^2}=-\frac{3}{2}\left(1+\frac{x^2}{2\alpha-1}-y^2\right).
\end{equation}
Integrating Eq. (\ref{hdoth2}) yields
\begin{equation}\label{hubble}
	H(N)=H_i \exp\left[-\frac{3}{2}\int_{N_i}^{N}\left(1+\frac{x^2}{2\alpha-1}-y^2\right){\rm d}N'\right],
\end{equation}
in which, $H_i$ is the Hubble parameter at the $e$-folds $N_i$.

Because of the physical condition $0\leqslant\Omega_\phi\leqslant1$, in an expanding universe with $H>0$, both the variables $x$ and $y$ must be in the range $(0,1)$ and satisfy the relation $x^2+y^2\leqslant 1$. For a positive potential, the parameter $\sigma$ ranges in the interval $(0,\infty)$. So, it is more appropriate to compact the phase space in this direction by defining a new dynamical variable $\zeta$ as \citep{Ng2001}
\begin{equation}\label{zeta}
	\zeta\equiv\frac{\sigma}{\sigma+1},
\end{equation}
which the range of its variation is (0, 1).
Using Eq. (\ref{zeta}) we rewrite the set of Eqs. (\ref{dyn1})-(\ref{dyn3}) in terms of the new variable $\zeta$ as follows
\begin{align}
	&\frac{{\rm d}x}{{\rm d}N}=f_1\equiv\frac{3}{2}\left(\frac{\zeta}{1-\zeta} y^2-x\frac{2\alpha}{2\alpha-1}\right)+\frac{3}{2}x\left(1-y^2+\frac{x^2}{2\alpha-1}\right),\label{dyn4}\\
	&\frac{{\rm d}y}{{\rm d}N}= f_2\equiv -\frac{3}{2}x y \frac{\zeta}{1-\zeta} +\frac{3}{2}y\left(1-y^2+\frac{x^2}{2\alpha-1}\right),\label{dyn5}\\
	&\frac{{\rm d}\zeta}{{\rm d}N}= f_3\equiv-3x\zeta^2(\Gamma-1)+\frac{3}{2}\zeta(1-\zeta)\frac{\alpha-1}{\alpha}\left(\frac{2\alpha}{2\alpha-1}-\frac{\zeta}{1-\zeta}\frac{y^2}{x}\right).\label{dyn6}
\end{align}

Now, the critical points of the autonomous system of Eqs. (\ref{dyn4})-(\ref{dyn6})  assuming a constant value for $\Gamma$ are the points in the phase space at which
\begin{equation}
	\frac{{\rm d}x}{{\rm d}N}=\frac{{\rm d}y}{{\rm d}N}=\frac{{\rm d}\zeta}{{\rm d}N}=0.
\end{equation}
Here, we use the linear stability theory in which the stability property of critical points is determined by evaluating eigenvalues of the Jacobian matrix, namely
\begin{equation}\label{Jacobian}
	J=  \left(
	\begin{array}{lll}
		\sfrac{\partial f_1}{\partial x}& \sfrac{\partial f_1}{\partial y} & \sfrac{\partial f_1}{\partial \zeta} \\
		\sfrac{\partial f_2}{\partial x} & \sfrac{\partial f_2}{\partial y} & \sfrac{\partial f_2}{\partial \zeta} \\
		\sfrac{\partial f_3}{\partial x} & \sfrac{\partial f_3}{\partial y} & \sfrac{\partial f_3}{\partial \zeta} \\
	\end{array}
	\right),
\end{equation}
at each critical point. When all eigenvalues of the Jacobian matrix are positive, the critical point is referred to as an unstable point, which repels the trajectories in the phase space. A critical point that has at least two eigenvalues with opposite signs is a saddle point, which attracts trajectories in some directions but repels them along
others. If all eigenvalues have negative values, the point will attract all nearby trajectories and is considered stable. For more details on the stability analysis of the critical points, see \citet{Bahamonde2018}. In Table \ref{cpoints1}, we list the critical points of the system along with their stability properties and the corresponding values of the parameters $\Omega_\phi$, $\omega_\phi$, and $\omega_{\rm eff}$. In the next, we describe the critical points as follows
\begin{table*}
	\centering
	\caption{\setlength{\baselineskip}{13pt}Existence and stability of the critical points as well as the corresponding parameters $\Omega_\phi$, $\omega_\phi$ and $\omega_{\rm eff}$.}
	\label{cpoints1}
	\begin{tabular}{cccccccccc} 
		\hline
		\hline
		Point& x & y & $\zeta$ & Existence &Eigenvalues& Stability & $\Omega_\phi$ & $\omega_\phi$ & $\omega_{\rm eff}$\\
		\hline
		$C_1$ & 1 & 0 & 0 &Always&\begin{tabular}{l}
			$e_1=3/(2\alpha-1)$\\
			$e_2=3(\alpha-1)/(2\alpha-1)$\\
			$e_3=3\alpha/(2\alpha-1)$
		\end{tabular}& Unstable & 1 & $\frac{1}{2\alpha-1}$ & $\frac{1}{2\alpha-1}$\\
		\hline
		$C_2$ & 0 & 0 & 0 &Always&\begin{tabular}{l}
			$e_1=3/2$\\
			$e_2=-3/(4\alpha-2)$\\
			$e_3=(\alpha-1)/(2\alpha-1)$
		\end{tabular}& Saddle & 0 & {\rm indeterminate} & 0 \\
		\hline
		$C_3$ & 0 & 1 & 0 &Always&\begin{tabular}{l}
			$e_1=3(\alpha-1)/(2\alpha-1)$\\
			$e_2=-3$\\
			$e_3=-3\alpha/(2\alpha-1)$
		\end{tabular}& Saddle & 1 & -1 & -1 \\
		\hline
		$C_4$ & 1 & 0 & 1 &For $\Gamma=1$&\begin{tabular}{l}
			$e_1=3/(2\alpha-1)$\\
			$e_2=-3(\alpha-1)/(2\alpha-1)$\\
			$e_3=\infty$
		\end{tabular}& Saddle & 1 & $\frac{1}{2\alpha-1}$ & $\frac{1}{2\alpha-1}$ \\
		\hline
		$C_5$ & 0 & 0 & 1 &Always&\begin{tabular}{l}
			$e_1=-3/(4\alpha-2)$\\
			$e_2=-3(\alpha-1)/(2\alpha-1)$\\
			$e_3={\rm indeterminate}$
		\end{tabular}& Stable & 0 & {\rm indeterminate} & 0\\
		\hline
		\hline
	\end{tabular}
\end{table*}
\begin{itemize}
	\item $C_1$: This critical point corresponds to a solution dominated by the kinetic part of the scalar field with $\Omega_\phi=1$, $\Omega_m=0$, $\omega_\phi=1/(2\alpha-1)$ and $\omega_{\rm eff}=1/(2\alpha-1)$. For $\alpha=1$, $\alpha=2$, and $\alpha\gg 1$, the effective EoS parameters are $\omega_{\rm eff}=1$, $\omega_{\rm eff}=1/3$, and $\omega_{\rm eff}\approx 0$, corresponding to an early-time solution of stiff matter, radiation domination, and matter domination, respectively. Eigenvalues of the Jacobian matrix at this point are $e_1=3/(2\alpha-1)$, $e_2=3(\alpha-1)/(2\alpha-1)$ and $e_3=3\alpha/(2\alpha-1)$. Hence, the point is unstable for $\alpha>1$.
	\item $C_2$: This point represents a matter dominated solution with $\Omega_\phi=0$, $\Omega_m=1$ and $\omega_{\rm eff}=0$. However, the EoS parameter of the scalar field is indeterminate at this point, meaning that its value depends on the initial conditions. It is a saddle point with eigenvalues $e_1=3/2$, $e_2=-3/(4\alpha-2)$ and $e_3=(\alpha-1)/(2\alpha-1)$.
	\item $C_3$: Here, the universe is in an accelerated regime dominated by the potential term of the scalar field. Eigenvalues of the Jacobian matrix are $e_1=3(\alpha-1)/(2\alpha-1)$, $e_2=-3$ and $e_3=-3\alpha/(2\alpha-1)$. So, the critical point is saddle for $\alpha>1$ and has properties of a de Sitter universe with $\Omega_\phi=1$, $\Omega_m=0$, $\omega_\phi=-1$ and $\omega_{\rm eff}=-1$.
	\item $C_4$: This critical point exists only for $\Gamma=1$. Note that at this point, $f_3=0$ but $f_1$ and $f_2$ are indeterminate due to the factor $y/(1-\zeta)$. Hence to ensure that $C_4$ is a critical point, we solve the system (\ref{dyn4})-(\ref{dyn6}) for $\Gamma=1$ with initial conditions in the vicinity of it, namely $(x\rightarrow 1, y\rightarrow 0, \zeta\rightarrow 1)$. The obtained results for various values of $\alpha>1$ show that $f_1(x\rightarrow 1,y\rightarrow 0, \zeta\rightarrow 1)\rightarrow 0$ and $f_2(x\rightarrow 1,y\rightarrow 0, \zeta\rightarrow 1)\rightarrow 0$ and as a result, the point $C_4$ is a critical point of the system. Same as the point $C_1$, this point is a solution dominated by the kinetic part of the scalar field. Eigenvalues are $e_1=3/(2\alpha-1)$, $e_2=-3(\alpha-1)/(2\alpha-1)$ and $e_3=\infty$. Hence, it is a saddle point located in the $\zeta=1$ plane corresponding with $\sigma=\infty$.
	\item $C_5$: At this point $f_1$, $f_2$ and $f_3$ are indeterminate. Hence, solving the system with initial conditions $(x\rightarrow 0, y\rightarrow 0, \zeta\rightarrow 1)$ for different values of $\alpha$ and $\Gamma$, we find that $f_1$, $f_2$ and $f_3$ tend to zero as $(x\rightarrow 0, y\rightarrow 0, \zeta\rightarrow 1)$. So, this point is a critical point corresponding with a matter-dominated solution with $\Omega_\phi=0$, $\Omega_m=1$, $\omega_{\rm eff}=0$, and $\omega_\phi$ is indeterminate. Eigenvalues of the Jacobian matrix at this point are $e_1=-3/(4\alpha-2)$, $e_2=-3(\alpha-1)/(2\alpha-1)$ and $e_3=(3/2)(1+x_c\zeta_c/(\zeta_c-1))$ where, $x_c=0$ and $\zeta_c=1$. It is clear that the eigenvalue $e_3$ is indeterminate due to the factor $x_c/(\zeta_c-1)$. Since, for $\alpha>1$, both the eigenvalues $e_1$ and $e_2$ have negative values, the stability of point $C_5$ depends on the sign of $e_3$, which is not determined. Here, we use a numerical method to investigate the stability of the critical point  $C_5$ by perturbing the system in the vicinity of it. This method has been quite successful in determining the stability characteristics of critical points \citep[see e.g.][]{Zonunmawia2017,Dutta2017,Dutta2019}. Figure \ref{fig:c5stability} shows the evolution of the phase space trajectories projected along $x$, $y$, and $\zeta$ axes, for $\Gamma=1$ and $\alpha=3$, assuming different initial conditions near the point $C_5$. It is clear from the figure that as $N\rightarrow \infty$, the trajectories approach the point $C_5$ with $x=0$, $y=0$, and $\zeta=1$. Similar behaviour is obtained for different values of $\alpha$ and $\Gamma$. Hence, this critical point is a stable node.
	
\end{itemize}
\begin{figure}
	\begin{tabular}{ccc}
		\includegraphics[width=0.33\columnwidth]{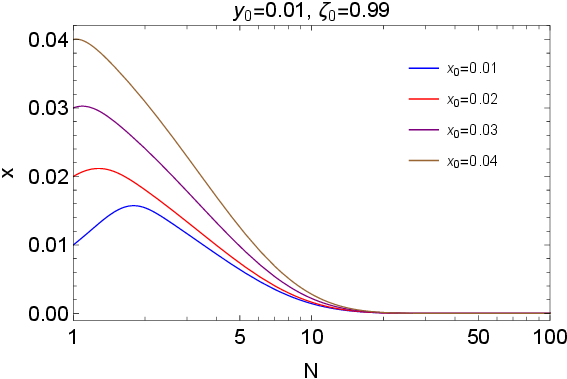}&
		\includegraphics[width=0.33\columnwidth]{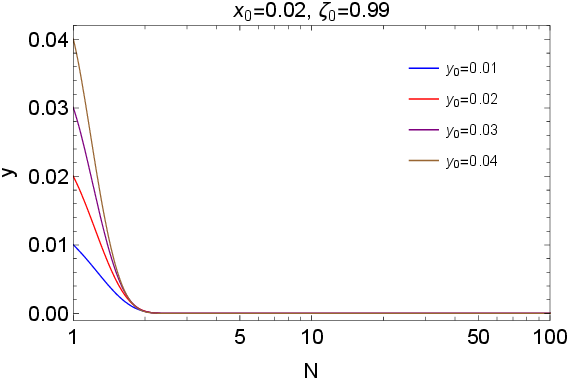}&
		\includegraphics[width=0.33\columnwidth]{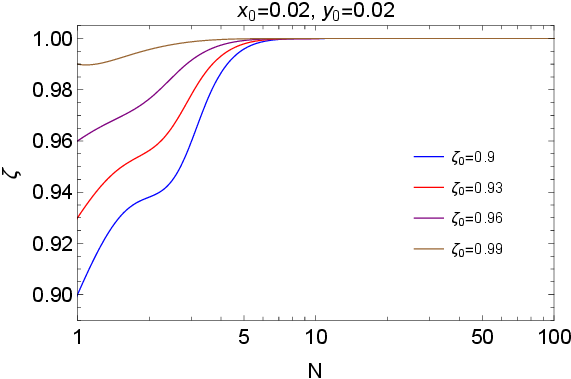}
	\end{tabular}
	\caption{\setlength{\baselineskip}{13pt}Projection of the evolution of phase space trajectories near the critical point $C_5$ along $x$-axis (left panel), $y$-axis (middle panel), and $\zeta$-axis (right panel) for $\Gamma=1$ and $\alpha=3$.}
	\label{fig:c5stability}
\end{figure}

\begin{figure}
	\begin{tabular}{cc}
		\includegraphics[width=0.45\columnwidth]{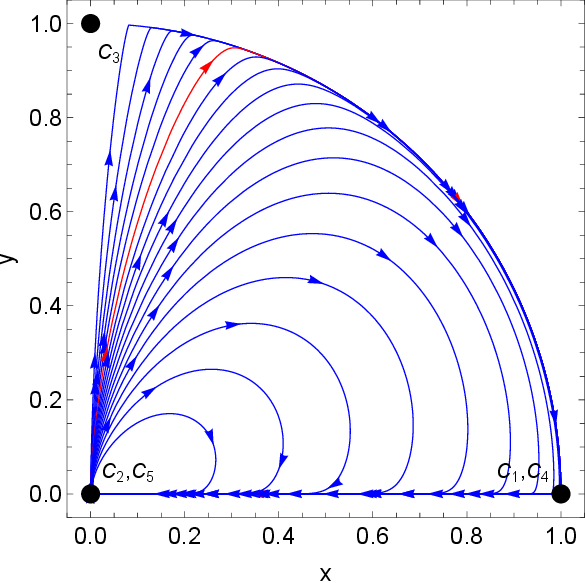}&
		\includegraphics[width=0.45\columnwidth]{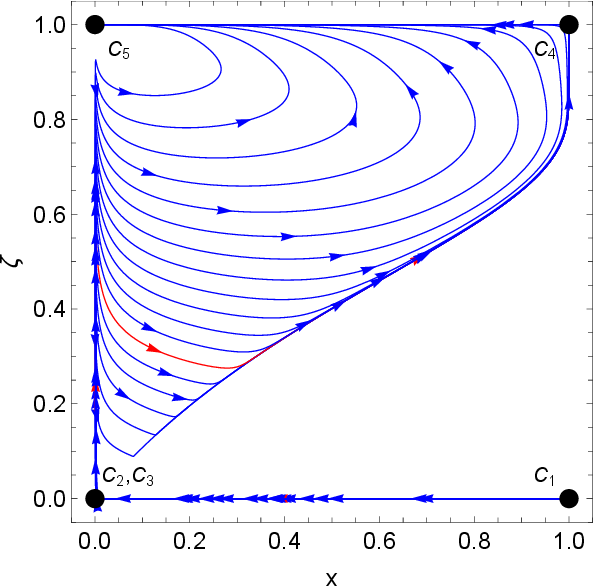}
	\end{tabular}
	\caption{\setlength{\baselineskip}{13pt}Projection of the phase space trajectories for $\Gamma=1$ and $\alpha=3$ along the $x-y$ plane (left panel) and $x-\zeta$ plane (right panel). Initial conditions are set as $x_0=0.001$, $\zeta_0=0.005$ with different values of $\zeta_0\in(0,1)$. The evolution directions are illustrated by arrows. The black dots denote the critical points.}
	\label{fig:phase_space}
\end{figure}

In addition, we solve the system of Eqs. (\ref{dyn4})-(\ref{dyn6}) considering the potential of the scalar field as follows
\begin{equation}\label{potential}
	V(\phi)=V_0\exp\left(-\lambda\frac{\phi}{M_P}\right),
\end{equation}
where $V_0$ and $\lambda$ are constant. Given that $\Gamma=1$ for this potential, the point $C_4$ acts as a critical point of the system.
We are interested in the solutions with a matter-dominated era in their evolution. Hence, we set the initial conditions as $x_0=0.001$, $y_0=0.005$, and different values of $\zeta_0\in(0,1)$.

In Fig. \ref{fig:phase_space}, the left and right panels illustrate the projection of the trajectories for $\alpha=3$ along the $x-y$ and $x-\zeta$ planes, respectively. All of the trajectories presented in the Figure evolve as $C_1\rightarrow C_2\rightarrow C_3 \rightarrow C_4 \rightarrow C_5$. For other values of $\alpha>1$, the overall shape of the orbits and their evolution direction would be similar to what is presented in Fig. \ref{fig:phase_space}. Therefore, based on the obtained results, the system starts its evolution from an unaccelerated scalar field-dominated point, $C_1$, and goes toward a matter-dominated universe described by the point $C_2$. Since the point $C_2$ is a saddle point, the orbits are repelled by it and continue their evolution toward the saddle point $C_3$, which is a universe dominated by the potential term of the scalar field. While the system evolves in the vicinity of the plane $x^2+y^2=1$ from $C_3$ toward the saddle critical point $C_4$, the universe remains scalar field dominated. However, through this evolution, the potential energy of the scalar field transforms to its kinetic energy. Ultimately, passing in the vicinity of the point $C_4$, the system evolves towards a stable universe dominated by matter as it approaches the point $C_5$.
\begin{figure}
	\centering
	\includegraphics[width=0.45\columnwidth]{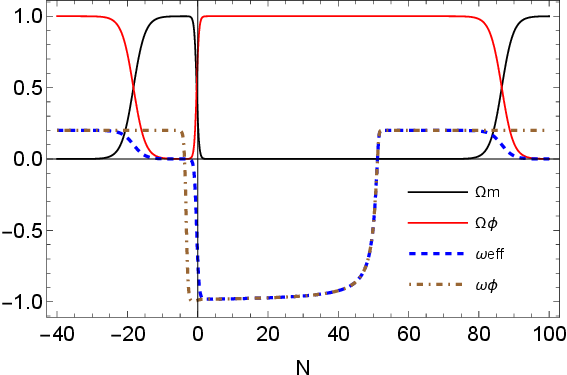}
	\caption{\setlength{\baselineskip}{13pt}Evolutions of $\Omega_m$ (black solid line), $\Omega_\phi$ (red solid line), $\omega_\phi$ (blue dashed line) and $\omega_{\rm eff}$ (purple dashed dotted line) for $\alpha=3$. Initial conditions are set as  $x_0=0.001$, $y_0=0.005$ and $\zeta_0=0.3$.}
	\label{fig:parameters_vs_N}
\end{figure}

The evolution of the system from $C_1$ to $C_5$ is more clear in Fig. \ref{fig:parameters_vs_N}, where the parameters $\Omega_m$, $\Omega_\phi$, $\omega_\phi$ and $\omega_{\rm eff}$ have been plotted versus $N$ for $\alpha=3$ assuming the initial conditions as $x_0=0.001$, $y_0=0.005$ and $\zeta_0=0.3$. In Fig. \ref{fig:phase_space}, the red trajectory corresponds to these initial conditions. Note that the dynamical system of Eqs. (\ref{dyn4})-(\ref{dyn6}) are invariant under translation $N\rightarrow N+N_0$. Hence, we are allowed to freely set the origin of the parameter $N$ on the phase space trajectories. In this context, we define the current time of the universe as the moment when $\Omega_\phi= 0.7$ and ${\rm d}\Omega_\phi/{\rm d}N>0$, assigning this specific point as $N=0$. Hence, the redshift $z$ is expressed in terms of the $e$-folds $N$ as $z=\exp(-N)-1$. From Fig. \ref{fig:parameters_vs_N}, it is evident that there exist three different points of equality of the matter and the scalar field as (i) one in the far past with $N\simeq -26$ corresponding with $z\simeq 5\times 10^{11}$. (ii) One in the near past, around $N\simeq 0.3$ or $z\simeq 0.35$. (iii) The last one in the far future at $N\simeq 43$ with $z\simeq -1$. Hence, at very high redshifts, there must have been a period of scalar field kinetic domination before the era of matter domination. However, this period occurs at very early times, when the effective description offered by the scalar field model is anticipated to break down due to the emergence of new physics, such as inflation. Due to this, the solutions for early times are typically disregarded in this model, and only the transition from matter to dark energy at late times is considered to be phenomenologically interesting. Also note that in the scalar field domination era for $N>0$, at the first stage the domination comes from the potential term of the scalar field during which $x^2\ll y^2$, $\omega_\phi<0$ and $\omega_{\rm eff}<0$. As the system evolves, the potential energy transforms to the non-canonical kinetic term of the scalar field and the universe becomes scalar field kinetic dominated with $x^2\gg y^2$, $\omega_\phi>0$ and $\omega_{\rm eff}>0$. At this transition point $x^2=y^2\simeq 0.5$, and from Eqs. (\ref{ophi}) and (\ref{oeff}) we have
\begin{equation}
	\omega_\phi=\omega_{\rm eff}\simeq \frac{1-\alpha}{2\alpha-1},
\end{equation}
where, for $\alpha=3$ we have $\omega_\phi=\omega_{\rm eff}=-0.4$ which takes place at $N\simeq 8$ equivalent to $z\simeq-0.999$. Results presented in Fig. \ref{fig:parameters_vs_N} clearly indicate that the model predicts a stable matter-dominated era as the eventual fate of the universe when $z\rightarrow -1$. However, justifying the prediction of the future and final fate of the universe using the model considered here is beyond the scope of this study since the main goal of the current paper is to investigate the formation of structures at the late-time matter to dark energy transition epoch. Notably, the evolution of the universe in this model is non-cyclic, as the future matter-dominated phase persists indefinitely.

\section{Parameters and evolution of the background}
The Lagrangian (\ref{lagrangian}) with the potential (\ref{potential}) has four parameters $\alpha$, $M$, $V_0$ and $\lambda$. While the parameters $M$, $V_0$, and $\lambda$ do not directly appear in the dynamical system (\ref{dyn4})-(\ref{dyn6}), their influence is implicitly encoded in the determination of the initial values of the dynamical variables. By utilizing Eqs. (\ref{dynvar}), (\ref{rhox}), and (\ref{potential}), we can express the dynamical variables $x$, $y$, and $\sigma$ in relation to both the model parameters and the quantities $H$, $\phi$, and $\phi'\equiv d\phi/d N=\dot{\phi}/H$ in the following manner
\begin{align}
	&x=\sqrt{\frac{2\alpha-1}{6}\tilde{\phi}'^{2\alpha}\left(\frac{\tilde{H}^2}{2\tilde{M}^4}\right)^{\alpha-1}},\label{xparam}\\
	&y=\frac{\sqrt{\tilde{V}_0\exp(-\lambda \tilde{\phi})}}{\sqrt{3}\tilde{H}},\label{yparam}\\
	&\sigma=\sqrt{\frac{2}{3}}\lambda\left(\frac{1}{2\alpha-1}\right)^\frac{1}{2\alpha}\left(\frac{\tilde{M}^4 y^2}{x^2 \tilde{V}_0 \exp(-\lambda \tilde{\phi})}\right)^\frac{\alpha-1}{2\alpha},
\end{align}
in which
\begin{equation}
	\tilde{H}\equiv \frac{H}{H_0},~~~ \tilde{\phi}\equiv\frac{\phi}{M_p},~~~\tilde{M}\equiv\frac{M}{\sqrt{M_p H_0}},~~~ \tilde{V}_0\equiv\frac{V_0}{M_p^2 H_0^2},
\end{equation}
with $H_0$ being the Hubble parameter at the present. Note that for $\alpha=1$, we have $\sigma=\sqrt{2/3}\lambda$ and the dynamical system (\ref{dyn1})-(\ref{dyn3}) reduces to that of \citet{Copeland1998}.

In order to justify the current acceleration of the universe, the slow-roll parameter $\eta_s\equiv M_P^2 V_{,\phi\phi}/V$ today needs to satisfy the condition $\eta_s \lesssim 1$, which is equivalent to $V_{,\phi\phi}\lesssim V/M_P^2\approx H_0^2$ \citep{Tsujikawa2013}. For the potential (\ref{potential}), this condition holds when $\lambda^2\lesssim 1$. For the potential energy today we have $V_0\exp(-\lambda\tilde{\phi}_0)\simeq H_0^2 M_P^2$ which for $\lambda\tilde{\phi}_0>0$ is equivalent to $\tilde{V}_0\gtrsim 1$. Also, from the Lagrangian (\ref{lagrangian}) we require that $M^4 \sim X \sim V$ and as a result $\tilde{M}=O(1)$.
\begin{figure*}
	\begin{tabular}{ll}
		(a)&(b)\\
		\includegraphics[width=0.4\columnwidth]{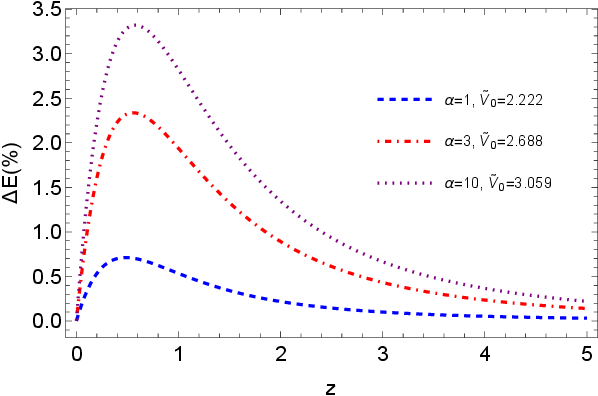} & \includegraphics[width=0.4\columnwidth]{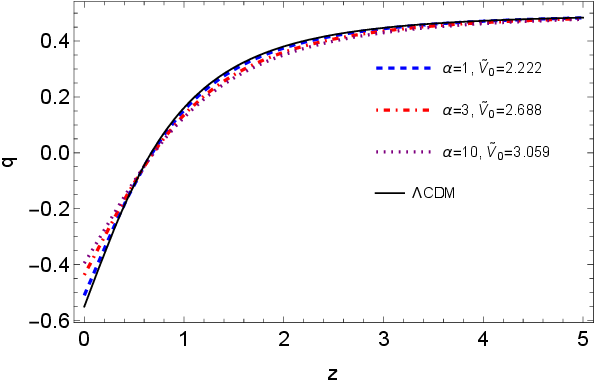}\\
		(c)&(d)\\
		\includegraphics[width=0.4\columnwidth]{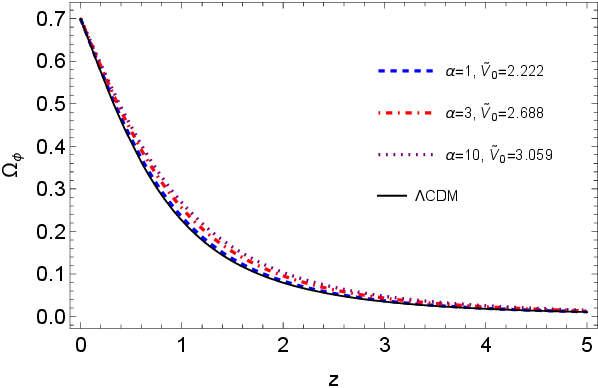} & \includegraphics[width=0.4\columnwidth]{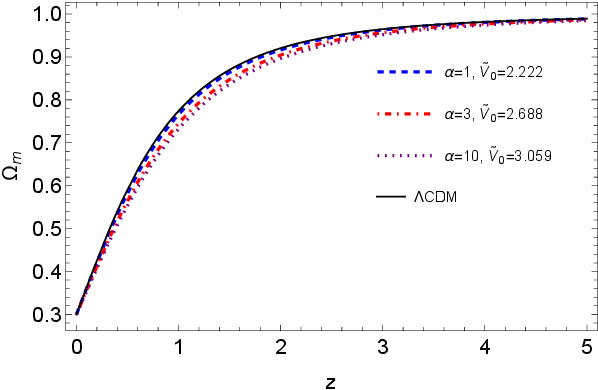}\\
		(e)&(f)\\
		\includegraphics[width=0.4\columnwidth]{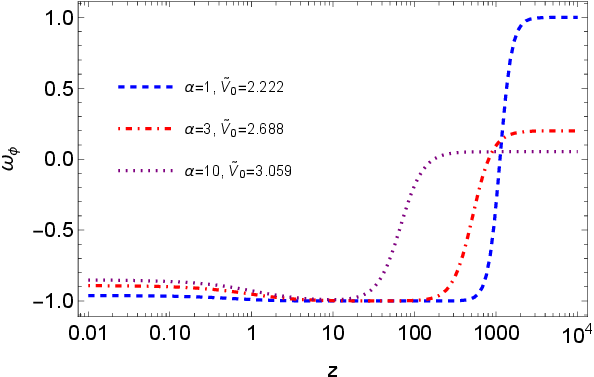} & \includegraphics[width=0.4\columnwidth]{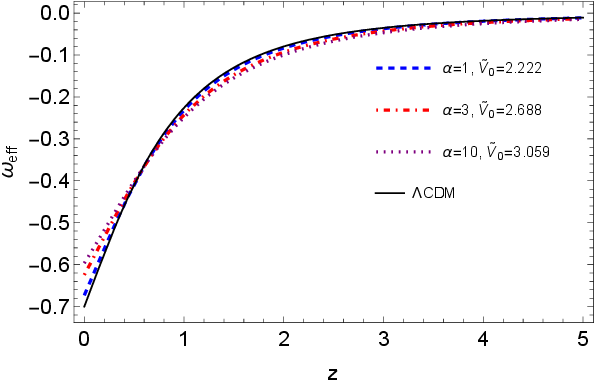}\\
		(g)&(h)\\
		\includegraphics[width=0.4\columnwidth]{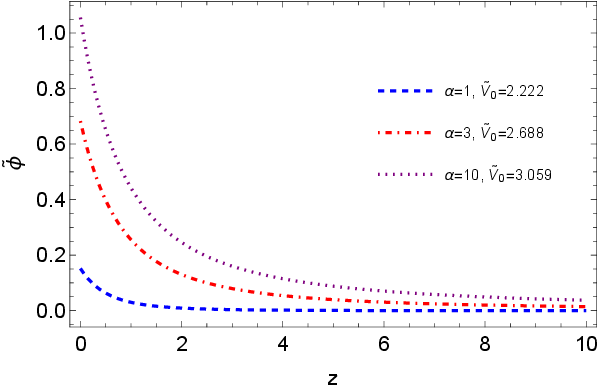} & \includegraphics[width=0.4\columnwidth]{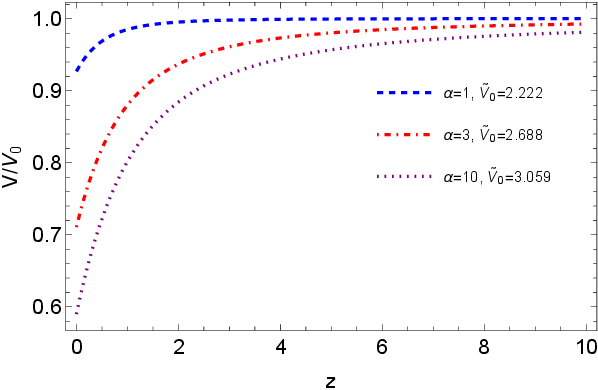}
	\end{tabular}
	\caption{\setlength{\baselineskip}{13pt}Evolution of the background quantities for $\lambda=0.5$ and $\tilde{M}=1$ in three different cases of  $\alpha=1$, $\tilde{V}_0=2.222$ (blue dashed line), $\alpha=3$, $\tilde{V}_0=2.688$ (red dashed-dotted line) and $\alpha=10$, $\tilde{V}_0=3.059$ (purple dotted line). (a): Relative difference of normalized Hubble parameter with that in $\Lambda$CDM model $\Delta E$, (b): The deceleration parameter $q$, (c): the density parameter of scalar field $\Omega_\phi$, (d): the density parameter of matter $\Omega_m$, (e): the EoS parameter of scalar field $\omega_\phi$, (f): the effective EoS parameter $\omega_{\rm eff}$, (g): the normalized scalar field $\tilde{\phi}$, (h): the normalized potential of the scalar field $V/V_0$. The solid lines in panels (c), (d), and (f) denote the results of $\Lambda$CDM.}
	\label{fig:background}
\end{figure*}

Here, we solve the dynamical system of Eqs. (\ref{dyn4})-(\ref{dyn6}) by setting the initial conditions at matter-radiation decoupling with $z_i=1100$. Setting $N=0$ for the present time, the redshift $z_i$ is equivalent to $N_i\simeq -7$. In order to obtain the initial values $x_i$, $y_i$ and $\zeta_i$, it is required to fix the quantities $H_i$, $\phi_i$ and $\phi '_i$ at $z=z_i$. 

The initial value of the Hubble parameter, $H_i$, is set to match the value obtained by the $\Lambda$CDM model assuming the present density parameters as $\Omega_{m_0}=0.3$ and $\Omega_{\Lambda_0}=0.7$, i.e. $\tilde{H}_i=\tilde{H}_{\rm{\Lambda CDM}}(z=1100)\simeq 2\times 10^4$. Note that, in principle, one only needs to ensure that $H=H_0$ today and the initial Hubble parameter $H_i$ can vary depending on the early-time dynamics of the model. In our case, we chose $H_i$ to match the value from the $\Lambda$CDM model at decoupling since it allows for a direct and meaningful comparison between the current model and the standard cosmological scenario. This alignment ensures that any deviations in the late-time expansion history can be attributed solely to the scalar field dynamics, rather than to differences in the initial conditions of the background evolution.

Since the initial conditions are set during the matter-dominated era, the value of $ \tilde{\phi}'_i $ is chosen to ensure that the scalar field remains subdominant, i.e., $ \Omega_{\phi,i} \ll 1 $. From Eq.~\eqref{Ophi}, this requirement translates into $ x_i^2 \ll 1 $ and $ y_i^2 \ll 1 $. As shown by Eq.~\eqref{yparam}, the condition $ y_i^2 \ll 1 $ is automatically satisfied for the range of parameters considered in this study. Substituting $ \tilde{H}_i = 2 \times 10^4 $ and $ \tilde{\phi}'_i = 10^{-4} $ into Eq.~\eqref{xparam}, we obtain:
$ x_i^2 = 1.7 \times 10^{-9}, \quad 3.3 \times 10^{-8}, \quad \text{and} \quad 1.6 \times 10^{-5}$
for $ \tilde{M} = 1 $ and $ \alpha = 1 $, 3, and 10, respectively. Increasing $ \tilde{\phi}'_i $ leads to larger values of $ x_i^2 $, which may eventually violate the condition $ \Omega_{\phi,i} \ll 1 $. Conversely, smaller values of $ \tilde{\phi}'_i $ are mathematically allowed, provided that numerical stability is preserved. In this work, we fix $ \tilde{\phi}'_i = 10^{-4} $ for all parameter sets considered.

For the initial value of the scalar field, $\tilde{\phi_i}$, we choose a small value to ensure that the field starts near the top of the potential, i.e., $V(\tilde{\phi_i}) \approx V_0$. In this work, we set $\tilde{\phi_i}= 10^{-3}$, which keeps the potential close to its maximum. Since $V_0$ is a free parameter, any constant shift in $\tilde{\phi_i}$ can be absorbed into a redefinition of $V_0$, making the choice of $\tilde{\phi_i}$ flexible without affecting the physical dynamics.

 By setting $\lambda=0.5$ and $\tilde{M}=1$, we solve Eqs. (\ref{dyn4})-(\ref{dyn6}) across varying $\alpha$ values. To ensure consistency with present-day parameters, we adjust the parameter $V_0$ so that $\Omega_\phi$ and $H$ match their current values, $\Omega_{\phi_0}=0.7$ and $\tilde{H}_0=1$, for each specific $\alpha$ value. Then, we use Eqs. (\ref{Ophi})-(\ref{dec}) and (\ref{hubble}) to obtain $\Omega_\phi$, $\Omega_m=1-\Omega_\phi$, $\omega_\phi$, $\omega_{\rm eff}$, $q$ and $H$ as a function of $z$. Also, using Eq. (\ref{potential}) and the definition of $y$ in Eq. (\ref{dynvar}) the scalar field $\phi$ and its potential $V$ are obtained as
\begin{align}
	&\tilde{\phi}=\frac{1}{\lambda}\ln\left(\frac{\tilde{V}_0}{3y^2 H^2}\right),\\
	&\tilde{V}=3y^2 H^2.
\end{align}

Figure (\ref{fig:background}) shows the results obtained for three different sets of parameters: i) $\alpha=1$, $\tilde{V}_0=2.222$; ii) $\alpha=3$, $\tilde{V}_0=2.688$; ii) $\alpha=10$, $\tilde{V}_0=3.059$. In panel (a) of the figure, the relative difference of the normalized Hubble parameters of the current model and $\Lambda$CDM, i.e., $\Delta E=100(H-H_{\rm{\Lambda CDM}})/H_{\rm{\Lambda CDM}}$ has been plotted versus redshift $z$. Results show that as the parameter $\alpha$ increases, the amount of $\Delta E$ increases. For $\alpha=1, 3$ and 10, the maximum value of the deviations are $\Delta E_{max}=0.71, 2.37$ and 3.32 percents which take place at $z=0.49, 0.55$ and 0.59, respectively. The evolution of the deceleration parameter $q$ has been shown in panel (b) of the figure. The transition redshift (where $q=0$) in all cases is almost the same as that predicted by the $\Lambda$CDM model around $z_t\simeq 0.67$. The deceleration parameter at the present time slightly differs from that of $\Lambda$CDM (here $q_0^{\rm {\Lambda CDM}}=-0.55$) with $q_0=-0.51, -0.44$ and $-0.39$ for $\alpha=1, 3$ and 10, respectively. Panels (c) and (d) of the figure illustrate the density parameters $\Omega_\phi$ and $\Omega_m$ versus $z$, respectively. It is clear from the figure that the density parameters evolve almost the same as those predicted by the $\Lambda$CDM model. Evolution of the equation of state of the scalar field $\omega_\phi$ has been plotted in panel (e) of the figure. It is evident that for all values of the parameter $\alpha$, the EoS parameter $\omega_{\phi}$ behaves like the quintessence model of DE, i.e $\omega_\phi>-1$. The present value of $\omega_{\phi}$ is obtained as $\omega_{\phi_0}=-0.96, -0.89$ and $-0.85$, respectively, for $\alpha=1, 3$ and 10. Panel (f) of the figure shows that the effective EoS parameter, $\omega_{\rm eff}$, evolves from a matter-dominated epoch with $\omega_{\rm eff}=0$ to the present value $\omega_{\rm eff_0}= -0.67, -0.62$ and $-0.60$, respectively for $\alpha=1, 3$ and 10. Variation of the normalized scalar field, $\tilde{\phi}$, and the corresponding potential, $V(\phi)$ normalized to $V_0$ has been illustrated, respectively, in panels (g) and (h) of Fig. \ref{fig:background}. Results show that, while the scalar field, $\tilde{\phi}$, evolves in an ascending manner, its potential decreases slowly during the evolution of the universe.

\section{Linear Perturbations}
\begin{figure}
	\centering
	\begin{tabular}{c}
		\includegraphics[width=0.45\columnwidth]{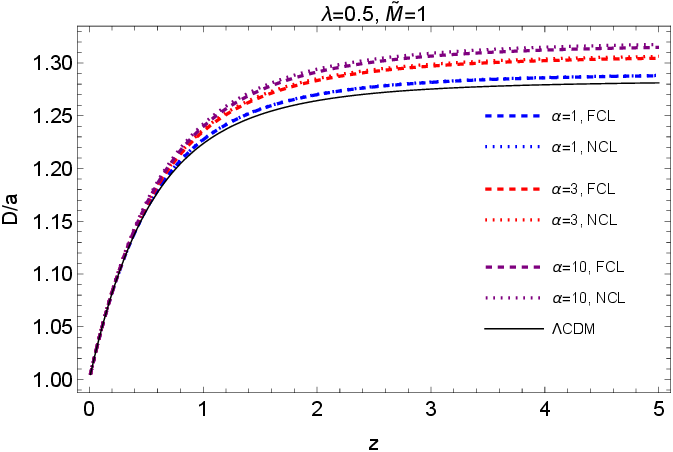}\\
		\includegraphics[width=0.45\columnwidth]{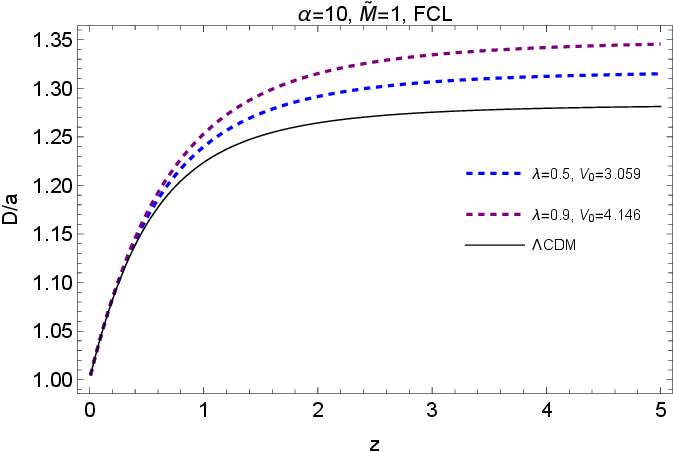}\\
		\includegraphics[width=0.45\columnwidth]{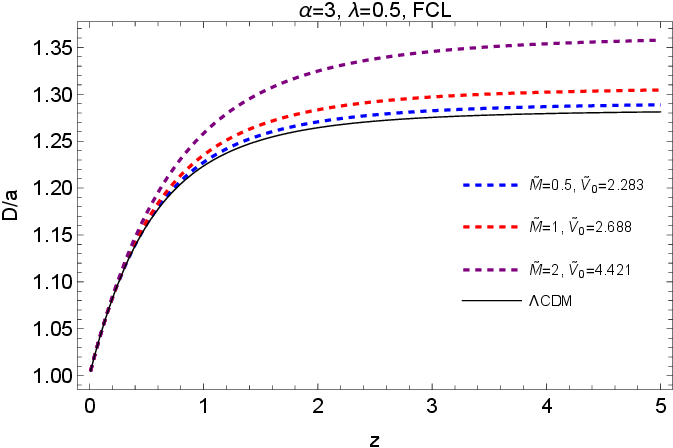}
	\end{tabular}
	\caption{\setlength{\baselineskip}{13pt}Evolution of the growth function $D\equiv \delta_m/\delta_{m_0}$ normalized by the scale factor, $a$ (the growth function of a pure matter model). The top, middle, and bottom panels display the variation of the function with respect to the parameters $\alpha$, $\lambda$, and $M$, respectively. In each panel, the solid curve represents the results of $\Lambda$CDM. }
	\label{fig:growth}
\end{figure}
In this section, we study the evolution of the perturbations of non-relativistic DM and DE in the linear regime. Following \cite{abramo2009}, in pseudo-Newtonian formalism, the equations governing the linear growth of the fluctuations on the sub-horizon scales ($k\gg H$) are as follows
\begin{align}
	&\dot{\delta}_m+\frac{\theta_m}{a}=0,\label{pert1}\\
	&\dot{\delta}_d+(1+\omega_d)\frac{\theta_d}{a}+3H(c_{\rm eff}^2-\omega_d)\delta_d=0,\label{pert2}\\
	&\dot{\theta}_m+H\theta_m-\frac{k^2}{a}\varphi=0,\label{pert3}\\
	&\dot{\theta}_d+H\theta_d-\frac{k^2 c_{\rm eff}^2\delta_d}{(1+\omega_d)a}-\frac{k^2}{a}\varphi=0,\label{pert4}
\end{align}
where $\delta_m\equiv\delta\rho_m/\rho_m$,  $\delta_d\equiv\delta\rho_d/\rho_d$, and $\theta_m\equiv\nabla\cdot v_m$, $\theta_d\equiv\nabla\cdot v_d$, are the density contrast and the divergence of the comoving peculiar velocity of DM and DE in the collapsing region, respectively. Also, $c_{\rm eff}^2\equiv\delta p_d/\delta\rho_d$ is the effective sound speed of DE and $\varphi$ is the perturbation of the Newtonian gravitational potential due to all the components, satisfying the Poisson equation in the Fourier space as follows \citep{abramo2009, fahimi2018}
\begin{equation}\label{poison}
	-\frac{k^2}{a^2}\varphi=\frac{3}{2}H^2\left(\Omega_m\delta_m+(1+3c_{\rm eff}^2)\Omega_d\delta_d\right).
\end{equation}
Applying this equation to eliminate $\varphi$ from Eqs. (\ref{pert3}) and (\ref{pert4}) and expressing the set of (\ref{pert1})-(\ref{pert4}) in terms of the redshift yields
\begin{align}
	&\delta'_m-\frac{\tilde{\theta}_m}{1+z}=0,\label{pert5}\\
	&(1+z)\delta'_d-3(c_{\rm eff}^2-\omega_d)\delta_d-(1+\omega_d)\tilde{\theta}_d=0,\label{pert6}\\
	&(1+z)\tilde{\theta}'_m-\left(2-(1+z)\frac{H'}{H}\right)\tilde{\theta}_m
	-\frac{3}{2}\left(\Omega_m\delta_m+(1+3c_{\rm eff}^2)\Omega_d\delta_d\right)=0,\label{pert7}\\
	&(1+z)\tilde{\theta}'_d-\left(2-(1+z)\frac{H'}{H}\right)\tilde{\theta}_d
	-\frac{3}{2}\left(\Omega_m\delta_m+(1+3c_{\rm eff}^2)\Omega_d\delta_d\right)+\frac{k^2 c_{\rm eff}^2 (1+z)^2}{H^2(1+\omega_d^2)}\delta_d=0,\label{pert8}
\end{align}
where $\tilde{\theta}\equiv\theta/(Ha)$ and the prime denotes a derivative with respect to the redshift, $z$. In general, $c_{\rm eff}$ is a function of time from the fact that the time dependence of the perturbations $\delta p_d$ and $\delta \rho_d$ are often different from each other. However, our focus here lies on examining the two extreme scenarios of $c_{\rm eff}=0$ and $c_{\rm eff}=1$. When $c_{\rm eff}=0$, the evolution equations of $\theta_m$ and $\theta_d$ (refer to Eqs. (\ref{pert7}) and (\ref{pert8})) become identical, indicating that the DE clusters in the same manner as DM, i.e. $\theta_m=\theta_d$. Following the terminology introduced by \cite{fahimi2018}, this corresponds to a scenario of full clustering (FCL) DE. Conversely, in the case of $c_{\rm eff}=1$, the DE is non-clustering (NCL), resulting in $\delta_d=\theta_d=0$.

Here, we numerically solve the set of Eqs. \eqref{pert5}-\eqref{pert8} for both the scenarios of full clustering and non-clustering DE. Initial conditions are set at a point in the matter domination epoch with $\Omega_m\simeq 1$ and $\Omega_d\simeq 0$. Solving Eqs. \eqref{pert5} and \eqref{pert7} at this point provides the initial conditions as
\begin{align}
	&\delta_{m_i}\simeq\frac{A}{1+z_i},\label{init1}\\
	&\tilde{\theta}_{m_i}\simeq-\delta_{m_i},\label{init2}
\end{align}
where $A$ is a constant. Note that, in the Einstein-de Sitter (EdS) universe, Eqs. \eqref{init1} and \eqref{init2} are exact solutions. In the special case of full clustering DE in which $c_{\rm eff}=0$ and $\tilde{\theta}_d=\tilde{\theta}_m$, the initial condition of the density contrast of DE is obtained from the solution of Eq. \eqref{pert6} as follows
\begin{equation}
	\delta_{d_i}\simeq\frac{1+\omega_{d_i}}{1-3\omega_{d_i}}\delta_{m_i}.\label{init3}
\end{equation}
Since the initial conditions \eqref{init1}-\eqref{init3} are directly proportional to the constant $A$, the solution of Eqs. \eqref{pert5}-\eqref{pert8} for both full clustering and non-clustering DE cases is linearly dependent on $A$. Therefore, in the linear regime, setting $A=1$ is permissible without sacrificing the general applicability of the solution. Here, we set the initial values of the perturbations at $z_i=10^3$ with $\delta_{m_i}=10^{-3}$ and solve the set of Eqs. \eqref{pert5}-\eqref{pert8} for both the FCL and NCL DE models.

Figure \ref{fig:growth} illustrates the evolution of the growth factor $D(z)\equiv \delta_m/\delta_m(z=0)$ divided by its value in an EdS model, $D=a$, for different sets of model parameters. As mentioned in the previous section, in each parameter set we adjust the value of $V_0$ to achieve $\Omega_{m_0}=0.3$ and $\tilde{H}=1$ with a desired level of accuracy. The top panel of the figure depicts the results obtained for $\lambda=0.5$, $\tilde{M}=1$, and three values of $\alpha=(1, 3, 10)$ in both the FCL and NCL scenarios. The results indicate that increasing the parameter $\alpha$ leads to an increase in the growth factor in both the FCL and NCL DE models. Furthermore, it is evident that for a given $\alpha$, the growth factor of the NCL DE is slightly larger than that of the FCL DE model. Moving on to the middle panel of Fig. \ref{fig:growth}, it demonstrates the dependency of the growth factor on the parameter $\lambda$ for $\alpha=10$ and $\tilde{M}=1$. Here, we observe that the growth factor increases as the parameter $\lambda$ increases. Finally, the bottom panel of the figure illustrates the dependency of the growth factor on the parameter $\tilde{M}$ for $\alpha=3$ and $\lambda=0.5$. We see that increasing the parameter $\tilde{M}$ results in an increase in the growth factor. Hence, the findings reveal that an increase in the parameters $\alpha$, $\lambda$, and $\tilde{M}$ is associated with a higher growth factor.

\section{Non-Linear growth of over-densities}
In this section, we investigate the growth of DE and DM over-densities in the non-linear regime. The transition from linear to non-linear evolution in cosmic structure formation is governed by the interplay between DM and DE fluctuations. In this context, the spherical collapse model (SCM) serves as an essential analytical tool for tracing the non-linear dynamics of overdense regions. This model describes the behaviour of a spherically symmetric perturbation that detaches from the homogeneous background due to its peculiar expansion rate. As self-gravity slows its expansion relative to the Hubble flow, the overdensity grows in contrast to the surrounding cosmic fluid. The system undergoes distinct evolutionary stages: i) turnaround (ta), where the collapsing region reaches its maximum radius, fully decouples from the background fluid, and begins to collapse independently; ii) virialization (vir), where it settles into a stable bound state. Crucially, the SCM assumes a top-hat density profile, ensuring that each fluid component remains homogeneous within the spherical region and that concentric shells within the over-density do not cross each other during collapse. Moreover, in the presence of DE, accelerated cosmic expansion suppresses large-scale gravitational potential growth while allowing dynamical DE to cluster and potentially form halos.

The equations governing the evolution of the perturbations in the SCM are as follows \citep{Hu1998, abramo2009, pace2014}
\begin{align}
	&\dot{\delta}_j=-3H(c_{{\rm eff}_j}^2-\omega_j)\delta_j-\left[1+\omega_j+(1+c_{{\rm eff}_j}^2)\delta_j\right]\frac{\theta}{a},\label{scm1}\\
	&\dot{\theta}=-H\theta-\frac{\theta^2}{3a}-4\pi G a\sum_{j}\rho_j\delta_j(1+3c_{{\rm eff}_j}^2),\label{scm2}
\end{align}
where the quantities $\delta_j$, $c_{{\rm eff}_j}^2$, and $\omega_j$ represent, respectively, the density contrast, the square of the effective sound speed, and the EoS parameter of component $j$. Each fluid component $j$ follows an individual equation of the form given in Eq. \eqref{scm1}, while Eq. \eqref{scm2} governs all components uniformly. This is a direct consequence of the top-hat density profile assumption in SCM and the fact that all fluids evolve in the same manner.
\begin{figure}
	\centering
	\begin{tabular}{c}
		\includegraphics[width=0.5\columnwidth]{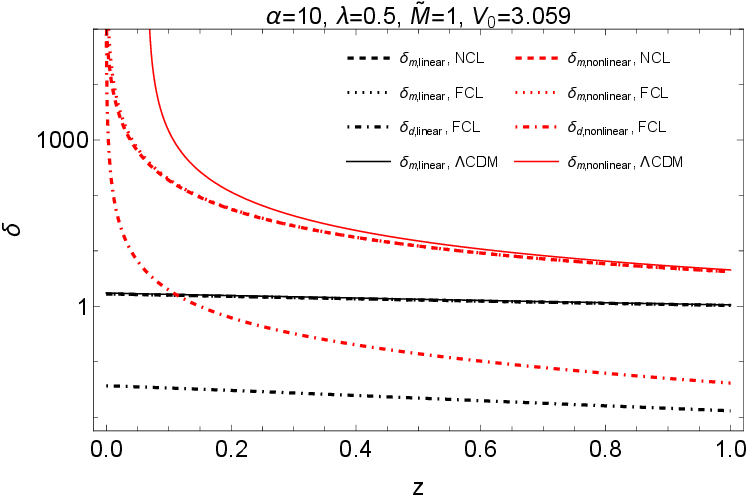}
	\end{tabular}
	\caption{\setlength{\baselineskip}{13pt}Evolution of DM and DE density contrasts in NCL and FCL DE models for $\alpha=10$, $\lambda=0.5$, $\tilde{M}=1$, and $V_0=3.059$. Initial conditions are chosen such that dark matter in the full clustering DE model collapses today. Black and red lines correspond to the linear and non-linear regimes, respectively.  Solid lines represent results from the $\Lambda$CDM model for comparison.}
	\label{fig:delta}
\end{figure}
The set of Eqs. \eqref{scm1} and \eqref{scm2} for DM and DE can be written in terms of the redshift $z$ as follows
\begin{align}
	&(1+z)\delta'_m-(1+\delta_m)\tilde{\theta}=0,\label{scm3}\\
	&(1+z)\delta'_d-3(c_{\rm eff}^2-\omega_d)\delta_d-\left[1+\omega_d+(1+c_{\rm eff}^2)\delta_d\right]\tilde{\theta}=0,\label{scm4}\\
	&(1+z)\tilde{\theta}'-\left(2-(1+z)\frac{H'}{H}\right)\tilde{\theta}-\frac{\tilde{\theta}^2}{3}
	-\frac{3}{2}\left[\Omega_m\delta_m+(1+3c_{\rm eff}^2)\Omega_d\delta_d\right]=0,\label{scm5}
\end{align} 
where the primes denote a derivative with respect to the redshift,$z$. We solve the system of Eqs. \eqref{scm3}-\eqref{scm5} for both the NCL and FCL DE models. In the NCL DE model, where $\delta_d = 0$, only Eqs. \eqref{scm3} and \eqref{scm5} need to be solved. Following \citet{pace2017}, we set the initial conditions at $z \sim 10^5$ to satisfy the EdS limit. At this redshift, the perturbations remain in the linear regime, so their initial values are given by Eqs. \eqref{init1}-\eqref{init3}. However, in contrast to the linear regime, the non-linear growth of over-densities is highly sensitive to the value of the constant $A$, which fine-tunes the initial perturbation values (see Eq. \eqref{init1}). Fig. (\ref{fig:delta}) shows the evolution of $\delta_m$ and $\delta_d$ in both the NCL and FCL DE models assuming the model parameters as $\alpha=10$, $\lambda=0.5$, $\tilde{M}=1$, and $V_0=3.059$. Initial conditions are chosen such that dark matter in the full clustering DE model collapses today, i.e., $\delta_{m_0}\longrightarrow \delta_\infty$. To satisfy the EdS limit, we set the collapse density contrast as $\delta_\infty=2\times 10^6$ \citep [see e.g.][]{Herrera2017,pace2017}. In Fig. \ref{fig:delta}, black and red curves denote the results obtained in linear and non-linear regimes, respectively. Results show that in the linear regime, the evolution of $\delta_m$ in both the NCL and FCL dark energy scenarios aligns closely with the predictions of the $\Lambda$CDM model. However, once the system transitions into the non-linear regime, the growth rate of dark matter fluctuations in the current model becomes noticeably slower than in $\Lambda$CDM. Notably, while the difference in the growth of DM perturbations between the FCL and NCL DE models is small, the FCL DE model exhibits a slight enhancement in perturbation growth.
\subsection{Parameters of spherical collapse}
In the context of the spherical collapse model, four key quantities play a central role in characterizing the nonlinear evolution of cosmic structures. i) The effective rate of expansion for the spherical region, $h(z)=H(1+\tilde{\theta}/3)$ \citep{abramo2009}. The turnaround redshift, $z_{ta}$, is the redshift at which the local expansion rate becomes zero, i.e., $h(z_{ta})=0$. ii) The critical density contrast $\delta_c\equiv\delta_{m, L}(z_c)$, which is the extrapolated value of the linear density contrast at $z_c$, serving as a bridge between linear and non-linear regimes. iii) The non-linear over-density at the turnaround redshift, $\zeta\equiv 1+\delta_{m, NL}(z_{ta})$.  iv) The virial overdensity, denoted by $\Delta_{\text{vir}}$, is defined as the ratio of the DM density within a structure to the background DM density: $\Delta_{\text{vir}} \equiv\rho_s/\rho_b = \zeta \left(a_c/a_{\text{ta}}\right)^3 \left(R_{\text{ta}}/R_c\right)^3$. Here, $R_{ta}$ and $R_c$ are the radius of the spherical region at turn around and collapse, respectively \citep[for more details see][]{Wang1998}. Together, these quantities provide essential input for halo modeling and cluster statistics. In the EdS universe, we have $\delta_c=1.686$, $\zeta=5.55$ and $\Delta_{vir}=178$ \citep[see][]{Tsujikawa2013}.
\begin{figure*}
	\centering
	\begin{tabular}{cc}
		\includegraphics[width=0.45\columnwidth]{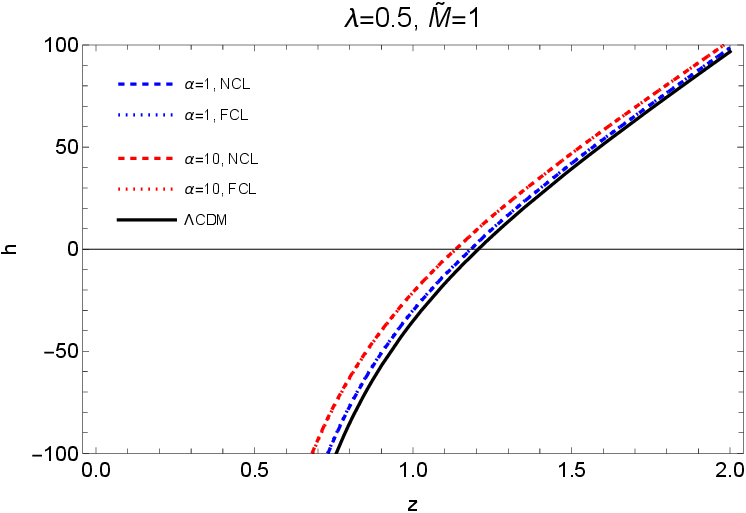} &\includegraphics[width=0.45\columnwidth]{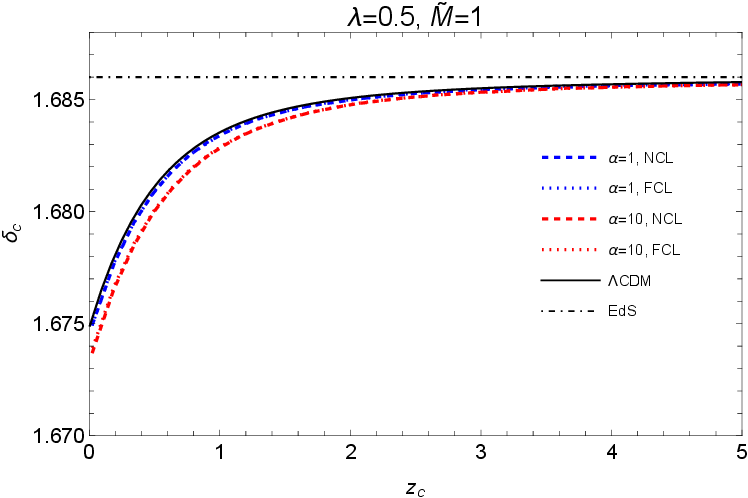}\\
		\includegraphics[width=0.45\columnwidth]{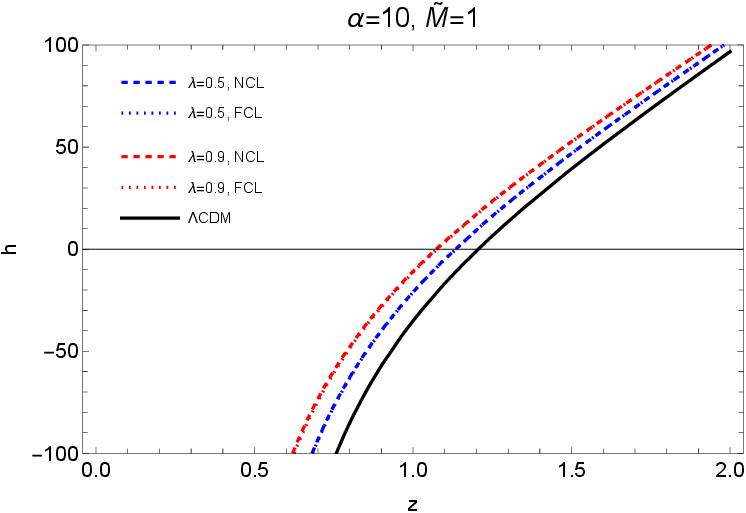} &\includegraphics[width=0.45\columnwidth]{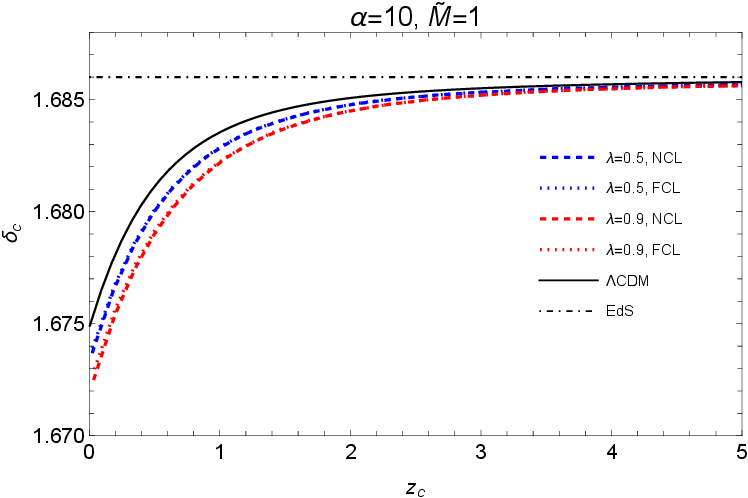}\\ \includegraphics[width=0.45\columnwidth]{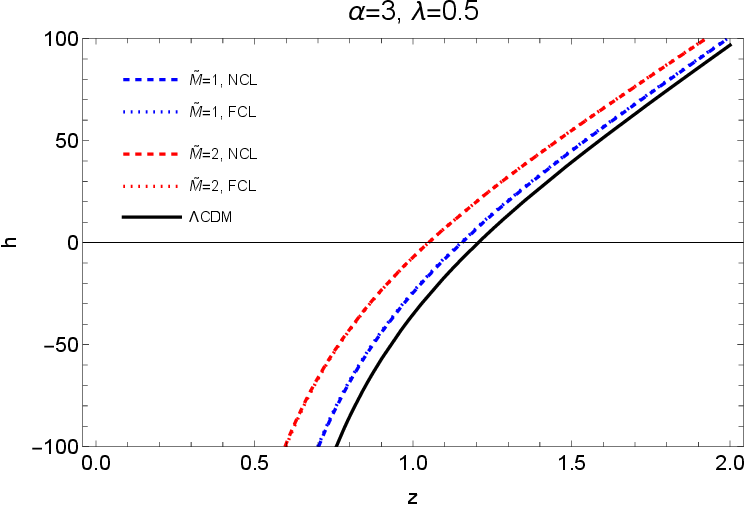}
		&\includegraphics[width=0.45\columnwidth]{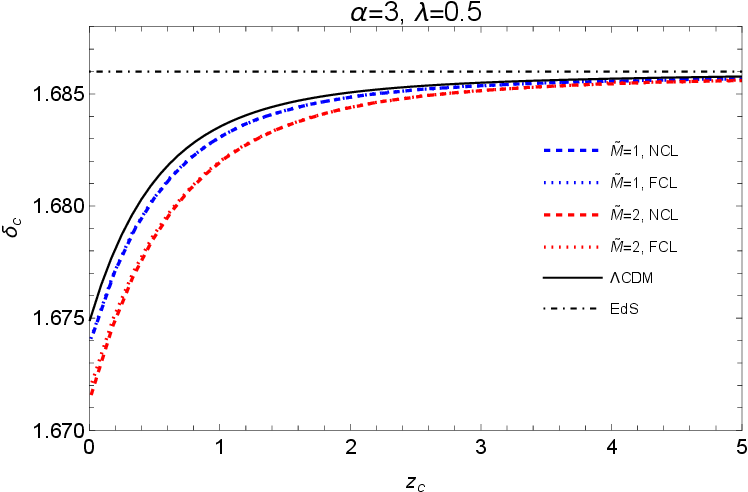}\\
	\end{tabular}
	\caption{\setlength{\baselineskip}{13pt}Evolution of the effective expansion rate, $h$ (left panels), and critical density contrast, $\delta_c$ (right panels) for different sets of the model parameters. Top, middle and bottom rows illustrate the variation of the results with respect to $\alpha$, $\lambda$ and $M$, respectively. The outcomes of the $\Lambda$CDM model is presented with solid curves for comparison.}
	\label{fig:parameter1}
\end{figure*}

\begin{figure*}
	\centering
	\begin{tabular}{cc}
		\includegraphics[width=0.45\columnwidth]{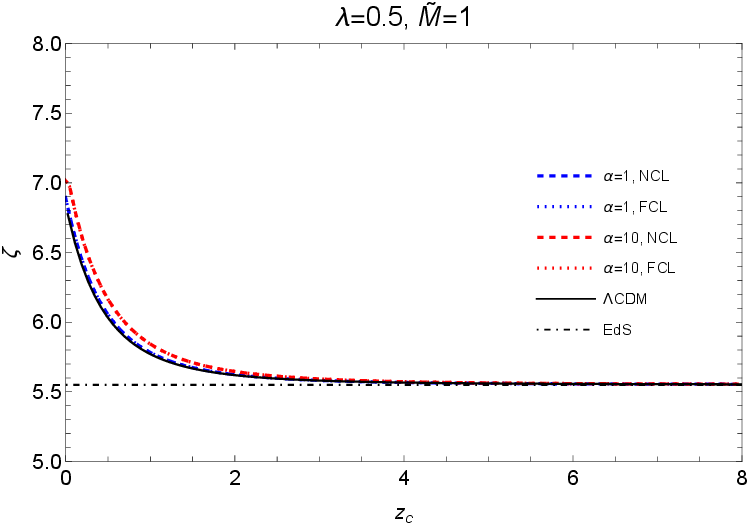} &\includegraphics[width=0.45\columnwidth]{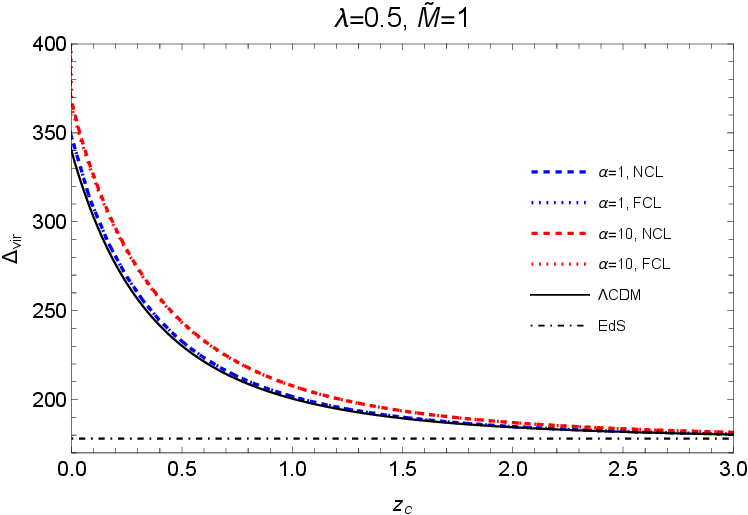}\\
		\includegraphics[width=0.45\columnwidth]{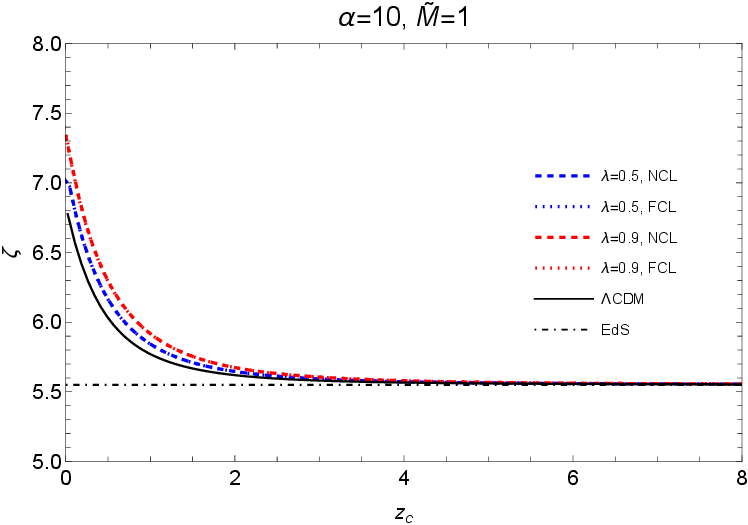} &\includegraphics[width=0.45\columnwidth]{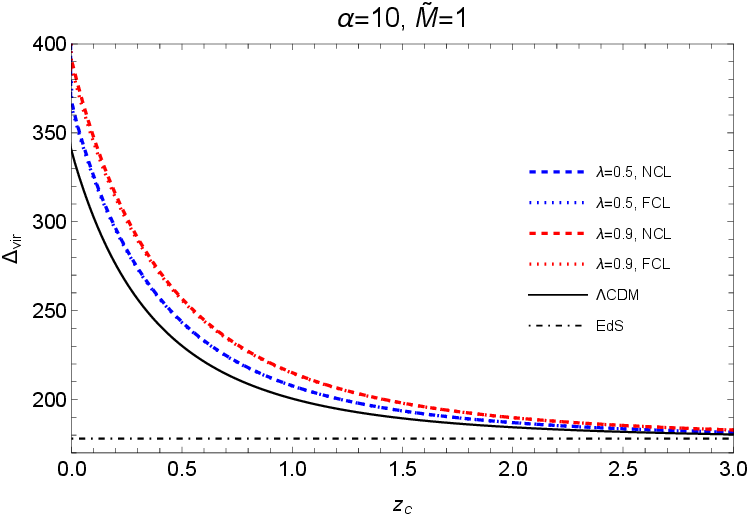}\\ \includegraphics[width=0.45\columnwidth]{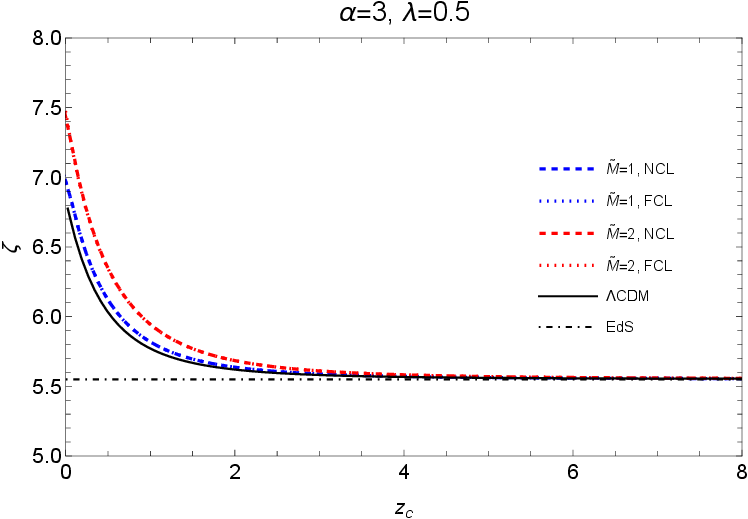}
		&\includegraphics[width=0.45\columnwidth]{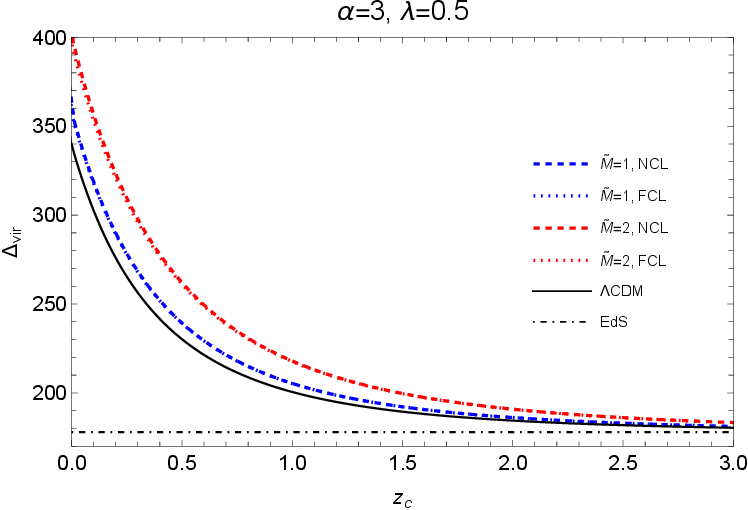}\\
	\end{tabular}
	\caption{\setlength{\baselineskip}{13pt}Same as Fig. \ref{fig:parameter1} but for the non-linear over-density at turn around, $\zeta$ (left panels), and the virial over-density, $\Delta_{vir}$ (right panels).}
	\label{fig:parameter2}
\end{figure*}
Figs. \ref{fig:parameter1} and \ref{fig:parameter2} illustrate the evolution of the quantities $h$, $\delta_c$, $\zeta$, and $\Delta_{vir}$ for different sets of model parameters in both the NCL (dashed curves) and FCL (dotted curves) DE models. However, the difference between these two DE scenarios is insignificant. For comparison, the results of the EdS and $\Lambda$CDM models are presented using dot-dashed and solid curves, respectively. In the figures, the top, middle, and bottom rows show the dependency of these quantities on the parameters $\alpha$, $\lambda$, and $M$, respectively. It's clear that as the model parameters increase, the results deviate further from those of $\Lambda$CDM. Across all parameter sets, the turnaround redshift, $z_{ta}$, is observed to be smaller than that of the standard model, as depicted in the left panels of Fig. \ref{fig:parameter1}. Note that the turnaround redshift depends on the initial values of the perturbation. It is clear that at high redshifts the quantities $\delta_c$, $\zeta$, and $\Delta_{vir}$ tend to their values in an EdS universe.

\subsection{The $f\sigma_8(z)$ function}

Here, we investigate the evolution of the function $f\sigma_8(z)\equiv f(z)\sigma_8(z)$ which encapsulates essential information about the growth of cosmic structures in the evolving universe and quantifies the strength of gravitational clustering. Here, the growth function $f(z)\equiv -(1+z){\delta}'_m/\delta_m$ represents the logarithmic derivative of the linear growth factor with respect to the scale factor and $\sigma_8(z)=D(z)\sigma_{8,0}$ in which $\sigma_{8,0}=\sigma_8(z=0)$ denotes the amplitude of matter density fluctuations on a scale of $8 h^{-1}$ Mpc. Following \citet{abramo2007}, we express $\sigma_{8,0}$ of the model in terms of its value in $\Lambda$CDM as follows
\begin{equation}
	\sigma_{8,0}^{\rm model}=\frac{\delta_c^{\rm model}(z=0)}{\delta_c^{\rm \Lambda CDM}(z=0)}\sigma_{8,0}^{\rm \Lambda CDM}.
\end{equation}
We apply the value of $\sigma_{8,0}=0.82$ as obtained in the $\Lambda$CDM model with $\Omega_{m_0}=0.3$ and $H_0=70~\rm km~s^{-1}~Mpc^{-1}$ \citep[for more details see section 4 of][]{fahimi2018}. Due to the small differences in $\delta_c$, the scaling factor $\delta_c^{\rm model}(z=0)/\delta_c^{\rm \Lambda CDM}(z=0)$ is nearly unity. Specifically, for the results of $\delta_c$ presented in the right column of Fig. \ref{fig:parameter1}, this ratio has values around $0.999$.

\begin{figure}
	\centering
	\begin{tabular}{c}
		\includegraphics[width=0.45\columnwidth]{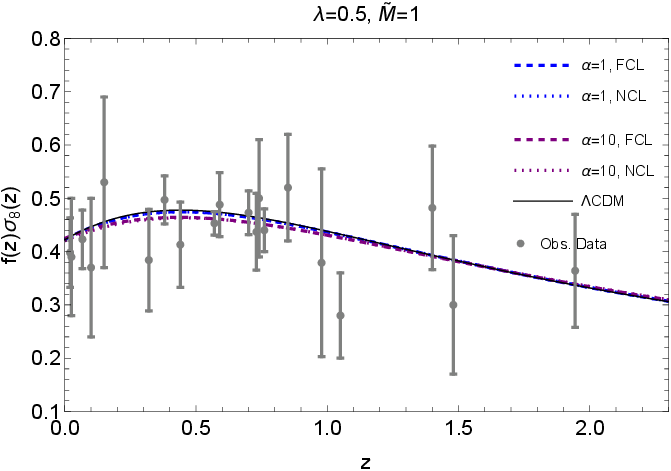}\\
		\includegraphics[width=0.45\columnwidth]{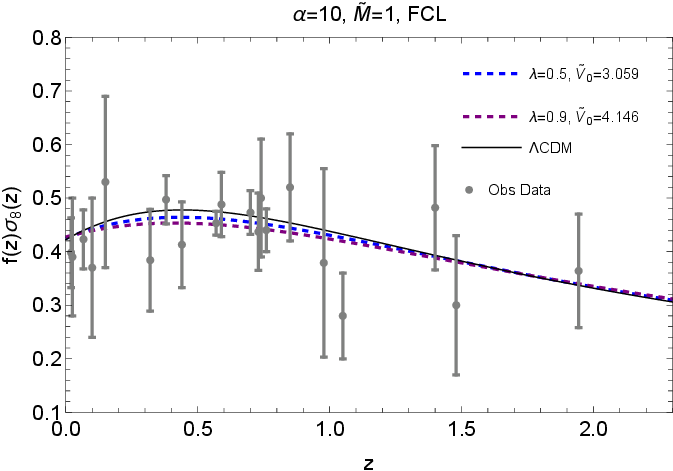}\\
		\includegraphics[width=0.45\columnwidth]{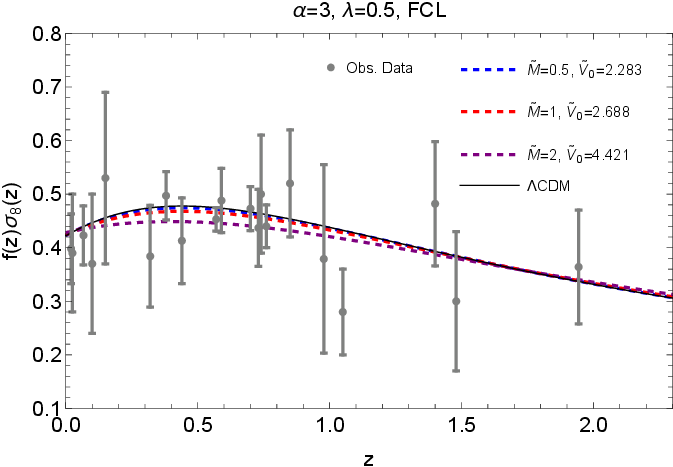}
	\end{tabular}
	\caption{\setlength{\baselineskip}{13pt}The function $f(z)\sigma_8(z)$ for different sets of the model parameters. Top, middle and bottom panels illustrate the variation of the results with respect to $\alpha$, $\lambda$ and $M$, respectively. Solid curve denotes the result of the $\Lambda$CDM model with $\Omega_{\Lambda_0}=0.7$. The gray dots represent the observational data compiled by \citet{avila2022}.}
	\label{fig:fsigma8}
\end{figure}

In Fig. \ref{fig:fsigma8}, we present the function $f(z)\sigma_8(z)$ derived from our model across various parameter sets. The solid black curve represents the outcome within the $\Lambda$CDM model with $\Omega_{\Lambda}=0.7$. The figure also shows the observational data points obtained according to the selection criteria outlined by \citet{avila2022}, marked as gray dots. These data points reflect direct measurements of $f\sigma_8$ obtained from uncorrelated and possibly correlated redshift bins, ensuring a comprehensive and robust data set for comparison and analysis. The top panel of Fig.  \ref{fig:fsigma8} illustrates the results obtained for $\lambda=0.5$, $\tilde{M}=1$ and two values of $\alpha=1 \& 10$ in both FCL and NCL DE scenarios. Considering a specific value of $\alpha$, it is evident that the disparity between the outcomes of FCL and NCL DE models is indiscernible. Furthermore, as the value of $\alpha$ grows, there is an increased deviation of the predicted $f(z)\sigma_8(z)$ in our model from that of the $\Lambda$CDM model. The middle and bottom panels of Fig. \ref{fig:fsigma8} show the variation of $f(z)\sigma_8(z)$ with respect to the parameters $\lambda$ and $\tilde{M}$, respectively. As mentioned previously, in each parameter set the value of $\tilde{V}_0$ is adjusted so that $\Omega_{\phi_0}=0.7$ and $\tilde{H}_0=1$. It is clear from the figure that increasing the parameter $\lambda$ or $\tilde{M}$ while other parameters are fixed, results in larger deviation from that obtained in the $\Lambda$CDM model.

\subsection{Halo number density}
Because the process of structure formation cannot be directly observed, it is practical to focus on a related observable quantity—namely, the comoving number density of virialized structures within a specific mass range. At redshift $z$, the comoving number density of virialized halos within the mass interval $m$ and $m+{\rm d}m$ is given by \citep{Press1974}
\begin{equation}
	\frac{{\rm d} n(m,z)}{{\rm d} m}=-\frac{\rho_{m_0}}{m}\frac{{\rm d}\ln \sigma(m,z)}{{\rm d}m}f(\sigma), \label{hnd}
\end{equation}
where $\rho_{m_0}$ denotes the present-day background matter density, and $\sigma$ represents the root-mean-square amplitude of mass fluctuations within spherical regions containing mass $m$. The mass function $f(\sigma)$ is described using the Sheth–Tormen formalism \citep{Sheth1999,Sheth2002}, namely
\begin{equation}
	f(\sigma)=A\sqrt{\frac{2a}{\pi}}\left[1+\left(\frac{\sigma^2(m,z)}{a\delta_c^2(z)}\right)^p\right]\frac{\delta_c(z)}{\sigma(m,z)} \exp\left(-\frac{a\delta_c^2(z)}{2\sigma^2(m,z)}\right),
\end{equation}
with parameter values $A= 0.3222$, $a = 0.707$, and $p= 0.3$. Following \citet{abramo2007}, the mass variance $\sigma(m,z)$ is related to its present-day value via
\begin{equation}
	\sigma(m,z)=D(z)\sigma_m.\label{sigmamz}
\end{equation}
The present-day variance, $\sigma_m$, of the smoothed linear matter density contrast is defined as
\begin{equation}
	\sigma_m^2=\int_{0}^{\infty}\frac{{\rm d}k}{k}\frac{k^3}{2\pi^2}P(k)W^2(kR),
\end{equation}
where, $R$ is the comoving scale enclosing a mass $m=(4\pi/3)R^3\rho_{m_0}$ and $W$ is the top-hat window function in Fourier space used for smoothing, defined as
\begin{equation}
	W(kR)=\frac{3}{(kR)^3}\Big(\sin(kR)-kR\cos(kR)\Big).
\end{equation}
The matter power spectrum $P(k)$, following \citet{Liddle1993} and \citet{Liddle1996}, is expressed as
\begin{equation}
	\frac{k^3}{2\pi^2}P(k)=\delta^2_{H_0}\left(\frac{ck}{H_0}\right)^{n_s+3}T^2(k),
\end{equation}
where $n_s=0.968$ \citep{Planck2016} is the primordial spectral index, $c$ is the speed of light, and $\delta_{H_0}$ is the present-day normalization of the spectrum. The transfer function $T(k)$ depends on cosmological parameters and the composition of matter in the Universe. Here we adopt the BBKS form \citep{Bardeen1986}:
\begin{equation}
		T(k)=\frac{\ln(1+2.34x)}{2.34 x}\left[1+3.89x+(16.1x)^2
		+(5.46x)^3+(6.71x)^4\right]^{-1/4}.
\end{equation} 
Here, $x\equiv k/(h\Gamma)$ in which the shape parameter $\Gamma$ \citep{Sugiyama1995} is given by
\begin{equation}
	\Gamma=\Omega_{m_0} h \exp\left(-\Omega_B-\frac{\Omega_B}{\Omega_{m_0}}\right),
\end{equation}
where the baryon density parameter is taken as $\Omega_B=0.016h^{-2}$ \citep{Copi1995a,Copi1995b}.

In scenarios where dark energy (DE) exhibits clustering, its perturbations can contribute to the total mass of dark matter (DM) halos. Therefore, it is essential to account for this contribution \citep{Creminelli2010, Basse2011, Batista2013, pace2014, Malekjani2015}. The ratio of DE mass to DM mass is expressed as $\epsilon(z)=m_d/m_m$. Assuming full DE clustering and a top-hat density profile, this ratio becomes:
\begin{equation}
	\epsilon(z)=\frac{\Omega_d}{\Omega_m}\left(\frac{\delta_d}{1+\delta_m}\right).
\end{equation}

\begin{figure}
	\centering
	\begin{tabular}{c}
		\includegraphics[width=0.45\columnwidth]{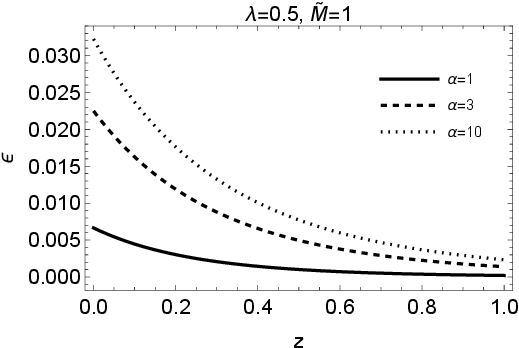}
	\end{tabular}
	\caption{\setlength{\baselineskip}{13pt}The ratio of scalar field DE mass to DM mass for $\lambda=0.5$, $\tilde{M}=1$ and three values of $\alpha=1$, 3 and 10.}
	\label{fig:epsilon}
\end{figure}
Fig. \ref{fig:epsilon} shows the evolution of $\epsilon(z)$ for $\lambda=0.5$, $\tilde{M}=1$ and three values of $\alpha=1$, 3 and 10. The figure shows that at early times, $\epsilon$ tends toward zero, implying that the contribution of dark energy mass to the total halo mass is negligible at high redshifts. Moreover, it is evident that for a fixed redshift $z$, the value of $\epsilon$ increases with increasing $\alpha$ or equivalently, with decreasing sound speed.

To account for the corrections associated with the dark energy mass, we replace $M$ in Eqs. \eqref{hnd} and \eqref{sigmamz} with $(1-\epsilon)M$. Then we use these modified equations to calculate the number density of halos with a mass greater than a specific mass $M$, namely
\begin{equation}
	n(>m)=\int_{m}^{\infty}\frac{{\rm d}n}{{\rm} m'}{\rm d}m'.
\end{equation}
Since the integration is performed numerically, we replace the upper limit with a sufficiently large mass --- here, $m=10^{18}$ --- beyond which the halo number density results converge to the desired level of accuracy.

\begin{figure}
	\centering
	\begin{tabular}{c}
		\includegraphics[width=0.45\columnwidth]{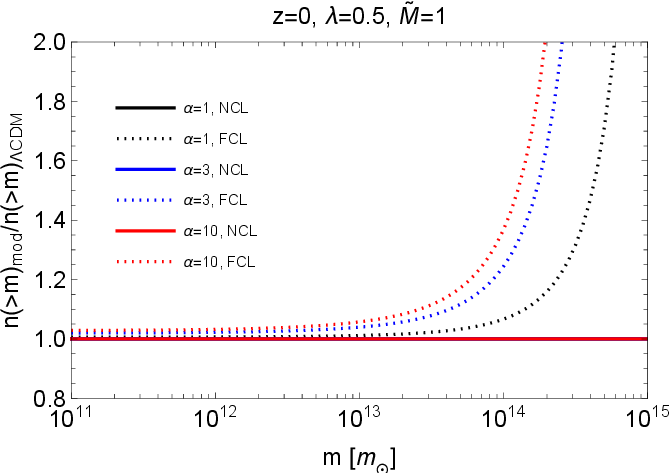}\\
		\includegraphics[width=0.45\columnwidth]{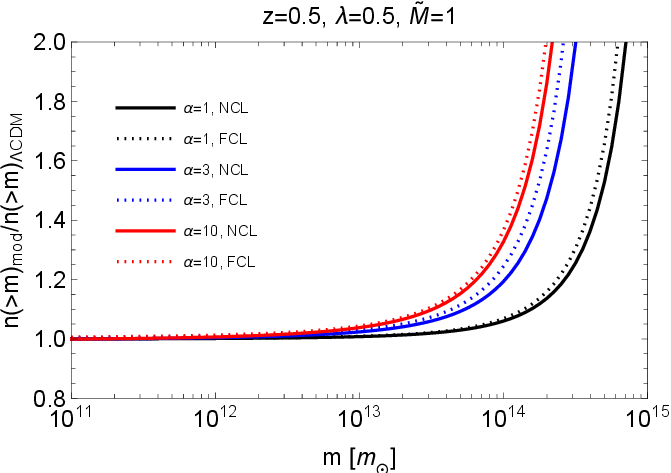} \\
		\includegraphics[width=0.45\columnwidth]{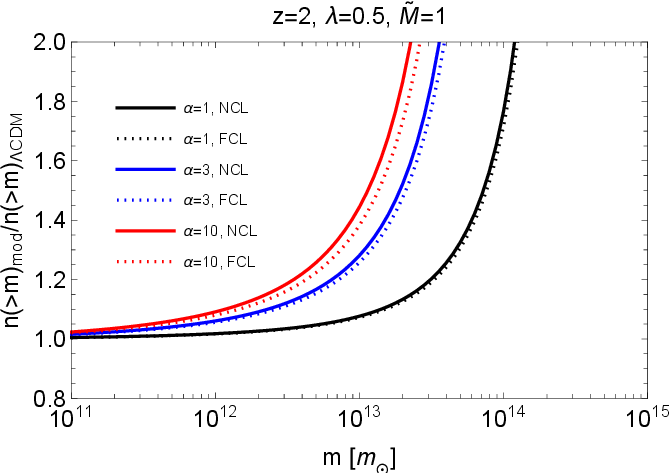}
	\end{tabular}
	\caption{\setlength{\baselineskip}{13pt}Relative number density of halo objects with mass exceeding $m$, computed for $\lambda=0.5$ and $\tilde{M}=1$, at redshifts $z=0$ (top panel), $z=0.5$ (middle panel), and $z=2$ (bottom panel). Black, blue, and red curves correspond to $\alpha=1$, 3 and 10, respectively. Solid lines represent the NCL DE model, while dotted lines indicate the FCL model.}
	\label{fig:hndz}
\end{figure}

\begin{figure}
	\centering
	\begin{tabular}{c}
		\includegraphics[width=0.5\columnwidth]{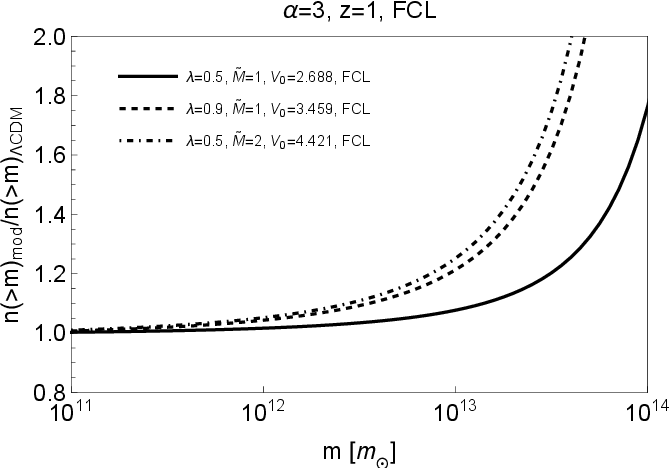}
	\end{tabular}
	\caption{\setlength{\baselineskip}{13pt}Relative number density of halo objects with mass exceeding $m$ in FCL DE scenario with $\alpha=3$ at redshift $z=1$ for three sets of parameters: i) $\lambda=0.5$, $\tilde{M}=1$ and $V_0=2.688$ (solid curve), ii) $\lambda=0.9$, $\tilde{M}=1$ and $V_0=3.459$ (dashed curve), iii) $\lambda=0.5$, $\tilde{M}=2$ and $V_0=4.421$ (dot-dashed curve).}
	\label{fig:hndalpha3}
\end{figure}

Figure \ref{fig:hndz} presents the relative number density of halos exceeding a given mass at fixed redshifts $z=0$, 0.5, and 2, for both the NCL and FCL models. At $z=0$, the outcomes of all non-clustering models closely align with those of the $\Lambda$CDM model. However, for clustering models with larger values of $\alpha$ (lower values of $C_s$), the number of objects exceeds that predicted by the $\Lambda$CDM model. The results indicate that at redshift $z=0.5$, the clustering model predicts a higher number of virialized halos compared to non-clustering model, whereas at $z=2$, it yields a lower halo count. Overall, the differences between clustering and homogeneous models, compared to the $\Lambda$CDM model, become more pronounced at higher redshifts and in the high-mass end of the halo mass function. This sensitivity arises from the Sheth–Tormen formalism \citep{Sheth1999,Sheth2002}, where the linear overdensity threshold $\delta_c$ plays a pivotal role. Even small variations in $\delta_c$ can significantly impact the abundance of massive halos \citep{Batista2013,pace2014,Heneka2017}. Figure \ref{fig:hndalpha3} illustrates how the relative halo number density in the clustering model with $\alpha=3$ at $z=1$ varies with changes in the parameters $\lambda$ and $M$. The figure reveals that increasing either $\lambda$ or $M$ leads to a higher halo count.

\section{Summary and Conclusions}

In this paper, we conducted a study of the growth of DM and DE perturbations in the context of a non-canonical scalar field model with an exponential potential. The model parameters are $\alpha$ and $M$, which appear in the kinetic term of the Lagrangian, as well as $V_0$ and $\lambda$, appearing in the potential term. Through the application of dynamical system analysis, we investigated the critical points and evolution of a spatially flat FLRW universe containing DE and pressureless DM. We found five critical points in the dynamical system, each representing distinct cosmological regimes. $C_1$ describes a universe dominated by the kinetic energy of the scalar field, exhibiting early-time behavior and instability. $C_2$ corresponds to a matter-dominated saddle point with an indeterminate scalar field equation of state. $C_3$ represents an accelerated expansion regime dominated by the scalar field potential, behaving as a saddle point. $C_4$ is another kinetic-dominated saddle point located on a specific hypersurface, characterized by one divergent eigenvalue. Lastly, $C_5$ is a matter-dominated solution with indeterminate stability analytically, but numerical analysis confirms it as a stable node. Overall, the cosmic evolution proceeds through these phases, transitioning from $C_1$ toward the stable configuration at $C_5$. Then, we examined the evolution of key background quantities, including the Hubble parameter $H$, the deceleration parameter $q$, the density parameters ($\Omega_m$, $\Omega_\phi$), the EoS of the scalar field $\omega_{\phi}$, the effective EoS $\omega_{\rm eff}$, the scalar field $\phi$ and its potential $V$. In each set of the parameters we adjusted $V_0$ so that we get $\Omega_\phi(z=0)=0.7$ and $\tilde{H}(z=0)=1$.

At the background level, the results showed that:
\begin{itemize}
	\item The relative difference in the normalized Hubble parameter between the current model and the standard $\Lambda$CDM model, denoted as $\Delta E$, becomes larger as the value of the parameter $\alpha$ is increased.
	\item For a fixed choice of the parameters $\lambda=0.5$ and $M=1$, the transition redshift (where $q = 0$) remains largely unchanged across different values of the parameter $\alpha$, closely matching the transition redshift predicted by the standard $\Lambda$CDM model. However, as the parameter $\alpha$ increases from 1 to 10, the deceleration parameter at present increases from $-0.51$ to $-0.39$. Also, the effective EoS parameter, $\omega_{\rm eff}$, evolves from a matter-dominated epoch with $\omega_{\rm eff}=0$ to the present value $\omega_{\rm eff_0}= -0.67$ and $-0.60$, respectively for $\alpha=1$ and 10.
	\item For all values of the parameter $\alpha$, the EoS parameter of the scalar field, $\omega_{\phi}$, behaves like the quintessence model of dark energy, where $\omega_{\phi}>-1$.
\end{itemize}

Within the linear perturbation regime, and employing the pseudo-Newtonian formalism, we derived the growth factor $D = \delta_m/\delta_{m_0}$, where $\delta_m$ denoted the matter density contrast and $\delta_{m_0}$ its present value. This growth factor was expressed relative to the case of a pure matter-dominated universe, in which the growth factor is simply $D=a$, with $a$
being the scale factor. We found that increasing the parameters $\alpha$, $\lambda$, and $\tilde{M}$ led to a higher growth factor compared to that obtained in the $\Lambda$CDM model.

We also examined the nonlinear evolution of perturbations by applying the spherical collapse model and derived four key parameters: the effective expansion rate of the collapsing structure, the critical density contrast, the nonlinear overdensity at turnaround, and the virial overdensity. These quantities were computed for both non-clustering and full clustering dark energy scenarios. Our analysis showed that the differences between these two cases in terms of spherical collapse parameters were negligible. However, as the model parameters increased, the deviations from the corresponding predictions of the $\Lambda$CDM model became more pronounced.

We calculated the function $f(z)\sigma_8(z)$ considering different sets of model parameters and compare the results with the outcome within the $\Lambda$CDM model. Results showed that for a given value of $\alpha$, the predicted $f\sigma_8$ in both cases of FCL and NCL DE scenarios are almost indistinguishable. Our findings showed that any increase in the value of the parameters $\alpha$, $\lambda$ and $M$ results to larger deviation from that obtained in the standard $\Lambda$CDM model.

Furthermore, the halo number density analysis demonstrates that clustering dark energy significantly affects the abundance of massive structures, especially at higher redshifts. While non-clustering models closely mimic $\Lambda$CDM at low redshift, clustering models predict enhanced halo formation at $z=0.5$ and suppressed counts at $z=2$. These deviations become more pronounced with increasing model parameters, offering potential observational signatures to distinguish between competing cosmological models. Overall, the results underscore the importance of dark energy clustering in shaping cosmic structure and provide a framework for testing scalar field cosmologies against large-scale structure data.

Although the results presented in the current study pertain to the specific case of an exponential potential, it becomes clear that, in general, the non-canonical scalar field model considered here has the potential to explain the universe's evolution from the matter-dominated epoch, both at the background level and in the growth of perturbations.

\subsection*{Acknowledgements}
The authors thank the anonymous referee for constructive comments.


\begin{thebibliography}{100}
	\bibitem[Riess {et al.} (1998)]{Riess1998}
	Riess A. G., Filippenko A. V., Challis P., et al. 1998, Astron. J., 116, 1009
	%
	\bibitem[Perlmutter {et al.} (1999)]{Perlmutter1999}
	Perlmutter S., Aldering G., Goldhaber G., et al. 1999, ApJ, 517,565
	%
	\bibitem[Kowalski {et al.} (2008)]{Kowalski2008}
	Kowalski M., Rubin D., Aldering G., et al. 2008, 686, 749
	%
	\bibitem[Goobar and Leibundgut (2011)]{Goobar2011}
	Goobar A., Leibundgut B. 2011, Annu. Rev. Nucl. Part. Sci., 61, 251
	%
	\bibitem[Komatsu {et al.} (2009)]{Komatsu2009}
	Komatsu E., Dunkley J.,  Nolta M. R., ApJS, 180, 330
	%
	\bibitem[Jarosik {et al.} (2011)]{Jarosik2011}
	Jarosik N., Bennett C. L., Dunkley J., et al. 2011, ApJS, 192, 15
	%
	 \bibitem[Komatsu (2011)]{Komatsu2011}
	 Komatsu E., 2011, ApJS, 192, 47
	 %
	 \bibitem[Planck Collaboration (2016)]{Planck2016}
	 Planck Collaboration XIII, 2016, A\&A, 594, A13
	 %
	 \bibitem[Planck Collaboration VI. (2020)]{Planck2020}
	 Planck Collaboration VI. 2020, A\&A, 641, A6
	 %
	 \bibitem[Cole {et al.} (2005)]{Cole2005}
	 Cole S., Percival W. J., Peacock J. A., et al. 2005, MNRAS, 362, 505
	 %
	 \bibitem[Tegmark {et al.} (2004)]{Tegmark2004}
	 Tegmark M., Strauss M. A., Blanton M. R.,  et al. 2004, Phys. Rev. D, 69, 103501
	 %
	 \bibitem[Eisenstein (2005)]{Eisenstein2005}
	 Eisenstein D. J. 2005, ApJ, 633, 560
	 %
	 \bibitem[Percival {et al.} (2010)]{Percival2010}
	 Percival W. J., Reid B. A., Eisenstein  D. J., et al. 2010, MNRAS, 401, 2148
	 %
	 \bibitem[Blake {et al.} (2011)]{Blake2011}
	 Blake C., Brough S., Colless M., et al. 2011, MNRAS, 415, 2876
	 %
	 \bibitem[Reid {et al.} (2012)]{Reid2012}
	 Reid B. A., Samushia, L., White, M., et al. 2012, MNRAS, 426, 2719
	 %
	 \bibitem[Alam {et al.} (2017)]{Alam2017}
	 Alam, S., Ata M., Bailey S., et al. 2017, MNRAS, 470, 2617
	 %
	 \bibitem[Alcaniz (2004)]{Alcaniz2004}
	 Alcaniz, J. S. 2004, Phys. Rev. D, 69, 83521
	 %
	 \bibitem[Wang and Steinhardt (1998)]{Wang1998}
	 Wang L., Steinhardt P. J. 1998, ApJ, 508, 483
	 %
	 \bibitem[Allen {et al.} (2004)]{Allen2004}
	 Allen S. W., Schmidt R. W., Ebeling H., Fabian A. C., Van Speybroeck L. 2004, MNRAS, 353, 457
	 %
	 \bibitem[Benjamin {et al.} (2007)]{Benjamin2007}
	 Benjamin J., Heymans C., Semboloni E., et al., 2007, MNRAS, 381, 702
	 %
	 \bibitem[Amendola {et al.} (2008)]{Amendola2008}
	 Amendola L., Kunz M., Sapone D., 2008, J. Cosmol. Astropart. Phys., 04, 13
	 %
	 \bibitem[Fu {et al.} (2008)]{Fu2008}
	 Fu L., Semboloni E., Hoekstra H., et al., 2008, A\&A, 479, 9
	 %
	 \bibitem[Shankaranarayanan and Johnson (2022)]{Shankaranarayanan2022}
	 Shankaranarayanan S., Johnson J.P. 2022, Gen. Relativ. Gravit., 54, 44
	 %
	 \bibitem[Peebles and Ratra (2003)]{Peebles2003}
	 Peebles P. J. E., Ratra B. 2003, Rev. Mod. Phys., 75, 559
	 %
	 \bibitem[Weinberg (1989)]{Weinberg1989}
	 Weinberg S. 1989, Rev. Mod. Phys., 61, 1
	 %
	 \bibitem[Sahni and Starobinsky (2000)]{Sahni2000}
	 Sahni V., Starobinsky A., 2000, Int. J. Mod. Phys. D, 9, 373
	 %
	 \bibitem[Carroll (2001)]{Carroll2001}
	 Carroll S. M. 2001, Liv. Rev. Relativ., 4, 56
	 %
	 \bibitem[Padmanabhan (2003)]{Padmanabhan2003}
	 Padmanabhan T. 2003, Phys. Rep., 380, 235
	 %
	\bibitem[Copeland {et al.}(2006)]{Copeland2006}
	Copeland E. J., Sami M., Tsujikawa S. 2006, Int. J. Mod. Phys. D, 15, 1753
	%
	\bibitem[Caldwell {et al.}(1998)]{Caldwell1998}
	Caldwell R. R., Dave R., Steinhardt P. J. 1998, Physical Review Letters, 80,
	1582
	%
	\bibitem[Erickson {et al.} (2002)]{Erickson2002}
	Erickson J. K., Caldwell R. R., Steinhardt P. J., Armendariz-Picon C.,
	Mukhanov V. 2002, Phys. Rev. Lett., 88, 121301
	%
	\bibitem[Caldwell (2002)]{Caldwell2002}
	Caldwell R. R. 2002, Phys. Lett. B, 545, 23
	%
	\bibitem[Caldwell {et al.}(2003)]{Caldwell2003}
	Caldwell R. R., Kamionkowski M., Weinberg N. N. 2003, Phys. Rev. Lett., 91, 71301
	%
	\bibitem[Elizalde {et al.} (2004)]{Elizalde2004}
	Elizalde E., Nojiri S., Odintsov S. D. 2004, Phys. Rev. D, 70, 43539
	%
	\bibitem[Armendariz-Picon {et al.} (2000)]{Armendariz2000}
	Armendariz-Picon C., Mukhanov V., Steinhardt P. J. 2000, Phys. Rev. Lett., 85, 4438
	%
	\bibitem[Armendariz-Picon {et al.} (2001)]{Armendariz2001}
	Armendariz-Picon C., Mukhanov V., Steinhardt P. J. 2001, Phys. Rev. D, 63, 103510
	%
	\bibitem[Chiba {et al.} (2000)]{Chiba2000}
	Chiba T., Okabe T., Yamaguchi M. 2000, Phys. Rev. D, 62, 23511
	%
	\bibitem[Kamenshchik {et al.} (2001)]{Kamenshchik2001}
	Kamenshchik A., Moschella U., Pasquier V. 2001, Phys. Lett. B, 511, 265
	%
	\bibitem[Bento {et al.} (2002)]{Bento2002}
	Bento M. C., Bertolami O., Sen A. A. 2002, Phys. Rev. D, 66, 43507
	%
	\bibitem[Gasperini {et al.} (2001)]{Gasperini2001}
	Gasperini M., Piazza F., Veneziano G. 2001, Phys. Rev. D, 65, 23508
	%
	\bibitem[Arkani-Hamed {et al.} (2004)]{Arkani2004}
	Arkani-Hamed N., Creminelli P., Mukohyama S., Zaldarriaga M. 2004, J. Cosmol. Astropart. Phys., 04, 1
	%
	\bibitem[Piazza and Tsujikawa (2004)]{Piazza2004}
	Piazza F., Tsujikawa S. 2004, J. Cosmol. Astropart. Phys., 7, 4
	%
	\bibitem[Cai (2007)]{Cai2007}
	Cai R.-G. 2007, Phys. Lett. B, 657, 228
	%
	\bibitem[Wei and Cai (2008)]{Wei2008}
	Wei H., Cai R.-G., 2008, Phys. Lett. B, 660, 113
	%
	\bibitem[Wei {et al.} (2005)]{Wei2005}
	Wei H., Cai R.-G., Zeng D.-F. 2005, Class. Quantum Gravity, 22, 3189
	%
	\bibitem[Koyama (2006)]{Koyama2006}
	Koyama K. 2006, J. Cosmol. Astropart. Phys., 3, 17
	%
	\bibitem[Malekjani {et al.} (2009)]{Malekjani2009}
	Malekjani M., Rahvar S., Haghi H. 2009, ApJ, 694, 1220
	%
	\bibitem[Brax and Valageas (2012)]{Brax2012}
	Brax P., Valageas P. 2012, Phys. Rev. D, 86, 63512
	%
	\bibitem[Nesseris (2013)]{Nesseris2013}
	Nesseris S. 2013, Phys. Rev. D, 88, 123003
	%
	\bibitem[Asadzadeh {et al.} (2016)]{Asadzadeh2016}
	Asadzadeh S., Khaledian M. S., Karami K. 2016, Iran. J. Astron. Astrophys., 3, 81
	%
	\bibitem[Nazari-Pooya {et al.} (2016)]{Nazari-Pooya2016}
	Nazari-Pooya N., Malekjani M., Pace F., Jassur D. M. Z., 2016, MNRAS, 458, 3795
	%
	\bibitem[Malekjani {et al.} (2017)]{Malekjani2017}
	Malekjani M., Basilakos S., Heidari N. 2017, MNRAS, 466, 3488
	%
	\bibitem[Nunes (2018)]{Nunes2018}
	Nunes R. C. 2018, J. Cosmol. Astropart. Phys., 5, 052
	%
	\bibitem[Linder and Jenkins (2003)]{Linder2003}
	Linder E. V., Jenkins A. 2003, MNRAS, 346, 573
	%
	\bibitem[Abramo {et al.} (2007)]{abramo2007}
	Abramo L. R., Batista R. C., Liberato L., Rosenfeld R. 2007, J. Cosmol. Astropart. Phys., 11, 12
	%
	\bibitem[Batista and Pace (2013)]{Batista2013}
	Batista R. C., Pace F. 2013, J. Cosmol. Astropart. Phys., 6, 44
	%
	\bibitem[Malekjani {et al.} (2015)]{Malekjani2015}
	Malekjani M., Naderi T., Pace F. 2015, MNRAS, 453, 4148
	%
	\bibitem[Naderi {et al.} (2015)]{Naderi2015}
	Naderi T., Malekjani M., Pace F. 2015, MNRAS, 447, 1873
	%
	\bibitem[Rezaei and Malekjani (2017)]{Rezaei2017}
	Rezaei M.,  Malekjani M. 2017, PRD, 96, 063519
	%
	\bibitem[Fahimi {et al.} (2018)]{fahimi2018}
	Fahimi K., Karami K., Asadzadeh S., Rezazadeh K. 2018, MNRAS, 481, 2393
	%
	\bibitem[Rezazadeh {et al.} (2020)]{Rezazadeh2020}
	Rezazadeh K., Asadzadeh S., Fahimi K.,Karami K., Mehrabi A. 2020, Annals of Physics, 422, 168299
	%
	\bibitem[Ziaie {et al.} (2020)]{Ziaie2020}
	Ziaie A.H., Moradpour H., Shabani H. 2020, Eur. Phys. J. Plus 135, 916
	%
	\bibitem[Guth (1981)]{Guth1981}
	Guth A. H. 1981, Phys. Rev. D, 23, 347
	%
	\bibitem[Linde (1990)]{Linde1990}
	Linde A. 1990, Phys. Lett. B, 238, 160
	%
	\bibitem[Mukhanov and Chibisov (1981)]{Mukhanov1981}
	Mukhanov V. F., Chibisov G. V. 1981, Soviet Journal of Experimental and Theoretical Physics Letters, 33, 532
	%
	\bibitem[Starobinsky (1982)]{Starobinsky1982}
	Starobinsky A. A. 1982, Physics Letters B, 117, 175.
	%
	\bibitem[Peebles (1993)]{Peebles1993}
	Peebles P. J. E. 1993, Principles of physical cosmology. Princeton Univ. Press, New Jersey
	%
	\bibitem[Peacock (1999)]{Peacock1999}
	Peacock J. A. 1999, Cosmological physics. Cambridge Univ. Press, Cambridge, p. 704
	%
	\bibitem[Abramo {et al.} (2008)]{abramo2008}
	Abramo L. R., Batista R. C., Liberato L., Rosenfeld R., 2008, Phys. Rev. D, 77, 67301
	%
	\bibitem[Gunn and Guth (1972)]{Gunn1972}
	Gunn J. E., Gott J. R. 1972, ApJ, 176, 1
	%
	\bibitem[Press and Schechter (1974)]{Press1974}
	Press W. H., Schechter P., 1974, ApJ, 187, 425
	%
	\bibitem[Bromm and Yoshida (2011)]{Bromm2011}
	Bromm V., Yoshida N. 2011, ARA\&A, 49, 373
	%
	\bibitem[Garriga and Mukhanov (1999)]{Garriga1999}
	Garriga J. Mukhanov V. F. 1999, Physics Letters B, 458, 219,
	%
	\bibitem[De~Santiago {et~al.}(2013)]{DeSantiago2013}
	De-Santiago J., Cervantes-Cota J. L., Wands D. 2013, Physical Review D, 87, 023502
	%
	\bibitem[Copeland {et al.}(1998)]{Copeland1998}
	Copeland E. J., Liddle A. R., Wands D. 1998, Physical Review D, 57, 4686
	%
	\bibitem[Zlatev {et al.} (1999)]{Zlatev1999}
	Zlatev I., Wang L., Steinhardt P. J. 1999, Physical Review Letters, 82, 896
	%
	\bibitem[Macorra and Piccinelli (2000)]{Macorra2000}
    Macorra A. d. l., Piccinelli G. 2000, Physical Review D, 61, 123503
	%
	\bibitem[Corasaniti and Copeland (2003)]{Corasaniti2003}
	Corasaniti P. S., Copeland E. J. 2003, Physical Review D, 67, 063521
	%
	\bibitem[Caldwell and Linder (2005)]{Caldwell2005}
	Caldwell R. R., Linder E. V. 2005, Physical Review Letters, 95, 141301
	%
	\bibitem[Linder (2006)]{Linder2006}
	Linder E. V. 2006, Physical Review D, 73, 063010
	%
	\bibitem[Yang {et al.} (2019)]{Yang2019}
	Yang W., Shahalam M., Pal B., Pan S., Wang A. 2019, Physical Review D,
	100, 023522
	%
	\bibitem[Joshi (2021)]{Joshi2021}
	Joshi T. 2021, in The 1st Electronic Conference on Universe. ECU 2021.
	MDPI, p. 39, doi:10.3390/ecu2021-09299
	%
	\bibitem[Ng {et al.} (2001)]{Ng2001}
	Ng S. C. C., Nunes N. J., Rosati F., Physical Review D, 64, 083510
	%
    \bibitem[Tsujikawa (2013)]{Tsujikawa2013}
    Tsujikawa S. 2013, Classical and Quantum Gravity, 30, 214003
	%
	\bibitem[Bahamonde {et al.} (2018)]{Bahamonde2018}
	Bahamonde S., Böhmer C. G., Carloni S., Copeland E. J., Fang W., Tamanini
	N. 2018, Physics Reports, 775, 1
	%
	\bibitem[Zonunmawia {et al.} (2017)]{Zonunmawia2017}
	Zonunmawia H., Khyllep W., Roy N., Dutta J., Tamanini N. 2017, Physical
	Review D, 96, 083527
	%
	\bibitem[Dutta {et al.} (2017)]{Dutta2017}
	Dutta J., Khyllep W., Tamanini N. 2017, Physical Review D, 95, 023515
	%
	\bibitem[Dutta {et al.} (2019)]{Dutta2019}
	Dutta J., Khyllep W., Zonunmawia H. 2019, The European Physical Journal
	C, 79
	%
	\bibitem[Abramo {et al.} (2009)]{abramo2009}
	Abramo L. R., Batista R. C., Liberato L., Rosenfeld R. 2009, Physical Review D, 79, 023516
	%
	\bibitem[Hu (1998)]{Hu1998}
	Hu W. 1998, ApJ, 506, 485
	%
	\bibitem[Pace {et al.} (2014)]{pace2014}
	Pace F., Batista R. C., Del Popolo A. 2014, MNRAS, 445, 648
	%
	\bibitem[Pace {et al.} (2017)]{pace2017}
	Pace F., Meyer S., Bartelmann M. 2017, J. Cosmol. Astropart. Phys., 10, 40
	%
	\bibitem[Herrera {et al.} (2017)]{Herrera2017}
	Herrera D., Waga I., Jor\'{a}s S.E. 2017,  Phys. Rev. D, 95, 064029
	%
	\bibitem[Avila {et al.} (2022)]{avila2022}
	Avila F., Bernui A., Bonilla A., Nunes R. C. 2022, The European Physical Journal
	C, 82, 594
	%
	\bibitem[Sheth and Tormen (1999)]{Sheth1999}
	Sheth R. K., Tormen G. 1999, MNRAS, 308, 119
	%
	\bibitem[Sheth and Tormen (2002)]{Sheth2002}
	Sheth R. K., Tormen G.  2002, MNRAS, 329, 61
	%
	\bibitem[Liddle and Lyth (1993)]{Liddle1993}
	Liddle A. R., Lyth D. H. 1993, Phys. Rep., 231, 1
	%
	\bibitem[Liddle {et al.} (1996)]{Liddle1996}
	Liddle A. R., Lyth D. H., Viana P. T. P., White M. 1996, MNRAS, 282, 281
	%
	\bibitem[Bardeen {et al.} (1986)]{Bardeen1986}
	Bardeen J. M., Bond J. R., Kaiser N., Szalay A. S. 1986, ApJ, 304, 15
	%
	\bibitem[Sugiyama (1995)]{Sugiyama1995}
	Sugiyama N. 1995, ApJS, 100, 281
	%
	\bibitem[Copi {et al.}(1995a)]{Copi1995a}
	Copi C. J., Schramm D. N., Turner M. S. 1995a, Phys. Rev. Lett., 75, 3981
	%
	\bibitem[Copi {et al.}(1995b)]{Copi1995b}
	Copi C. J., Schramm D. N., Turner M. S. 1995b, Science, 267, 192
	%
	\bibitem[Creminelli {et al.}(2010)]{Creminelli2010}
	Creminelli P., D’Amico G., Norena J., Senatore L., Vernizzi F., 2010, J. Cosmol. Astropart. Phys., 3, 27
	%
	\bibitem[Basse {et al.}(2011)]{Basse2011}
	Basse T., Bjælde O. E., Wong Y. Y. Y., 2011, J. Cosmol. Astropart. Phys., 10, 38
	%
	\bibitem[Heneka {et al.} (2017)]{Heneka2017}
	Heneka C., Rapetti D., Cataneo M., Mantz A. B., Allen S. W., von der Linden A. 2017, MNRAS, 473, 3882
	%
\end{thebibliography}
\end{document}